\documentclass[draft]{agujournal2019}
\usepackage{url} %this package should fix any errors with URLs in refs.
\usepackage{xurl}
\usepackage{lineno}
\usepackage{soul}
\usepackage{gensymb}
\usepackage{diagbox}
\usepackage{graphicx}
\usepackage{tikz}

\usepackage{ragged2e} % alignment adjustments with improved line breaks.
\justifying % aligns both the left and right margins of the text block

\draftfalse

\journalname{JGR: Machine Learning and Computation}

\begin{document}
%% ------------------------------------------------------------------------ 
%
%                               Title
%
%% ------------------------------------------------------------------------ 
%TC:ignore
\title{DispFormer: A Pretrained Transformer Incorporating Physical Constraints for Dispersion Curve Inversion}
%TC:endignore
%% ------------------------------------------------------------------------ 
%
%                       Author & Affiliation
%
%% ------------------------------------------------------------------------ 
%TC:ignore
\authors{Feng Liu\affil{1,2}, 
        Bao Deng\affil{3}, 
        Rui Su\affil{2*}, 
        Lei Bai\affil{2},
        Wanli Ouyang\affil{2},
        }

\affiliation{1}{School of Electronic Information and Electrical Engineering, Shanghai Jiao Tong University, Shanghai 200240, China}
\affiliation{2}{Shanghai Artificial Intelligence Laboratory, Shanghai 200232, China}
\affiliation{3}{Laboratory of Seismology and Physics of Earth’s Interior, School of Earth and Space Sciences, University of Science and Technology of China, 96 Jinzhai Road, Hefei 230026 Anhui, China}

\correspondingauthor{Rui Su}{surui@pjlab.org.cn}
%TC:endignore
%% ------------------------------------------------------------------------ 
%
%                               Keypoints
%
%% ------------------------------------------------------------------------ 
%TC:ignore
%% Keypoints, final entry on title page.
\begin{keypoints}
\item DispFormer integrates Transformer and depth-sensitive wave physics for accurate S-wave inversion from variable-length dispersion data.

\item Zero-shot DispFormer generates reliable profiles without labels, while few-shot learning outperforms traditional methods with limited data.

\item Synthetic and real-world studies show robust handling of noisy, incomplete dispersion data, improving inversion accuracy and efficiency.

\end{keypoints}
%TC:endignore
%% ------------------------------------------------------------------------ 
%
%                        Abstract & PLS
%
%% ------------------------------------------------------------------------ 
\begin{abstract}

    Surface wave dispersion curve inversion is crucial for estimating subsurface shear-wave velocity ($v_s$), yet traditional methods often face challenges related to computational cost, non-uniqueness, and sensitivity to initial models. While deep learning approaches show promise, many require large labeled datasets and struggle with real-world datasets, which often exhibit varying period ranges, missing values, and low signal-to-noise ratios. To address these limitations, this study introduces DispFormer, a transformer-based neural network for $v_s$ profile inversion from Rayleigh-wave phase and group dispersion curves. DispFormer processes dispersion data independently at each period, allowing it to handle varying lengths without requiring network modifications or strict alignment between training and testing datasets. A depth-aware training strategy is also introduced, incorporating physical constraints derived from the depth sensitivity of dispersion data. DispFormer is pre-trained on a global synthetic dataset and evaluated on two regional synthetic datasets using zero-shot and few-shot strategies. Results show that even without labeled data, the zero-shot DispFormer generates inversion profiles that outperform the interpolated reference model used as the pretraining target, providing a deployable initial model generator to assist traditional workflows. When partial labeled data available, the few-shot trained DispFormer surpasses traditional global search methods. Real-world tests further confirm that DispFormer generalizes well to dispersion data with varying lengths and achieves lower data residuals than reference models. These findings underscore the potential of DispFormer as a foundation model for dispersion curve inversion and demonstrate the advantages of integrating physics-informed deep learning into geophysical applications.

\end{abstract}

%TC:ignore
\section*{Plain Language Summary}

    Understanding Earth's subsurface structure is critical for earthquake hazard assessment, geological studies, and resource exploration. Scientists analyze surface wave dispersion curves, which describe how seismic waves travel at different frequencies, to infer underground rock properties. However, traditional computation methods struggle to balance accuracy and efficiency. While deep learning offers promising alternatives, early methods rely on large labeled datasets and struggle with real-world data, which is often incomplete and noisy. To address these challenges, this study introduces DispFormer, a deep learning framework inspired by language-processing algorithms. DispFormer independently processes each point in dispersion curves while incorporating geological principles, enabling it to handle curves of varying lengths and quality. It is first trained on a global synthetic dataset and later validated on both synthetic and real-world data. Synthetic results show that, even without regional training data, DispFormer generates more accurate initial subsurface models than standard interpolation methods, thereby better assisting traditional inversion workflows. Moreover, when partial labeled data is available, it outperforms traditional methods in terms of both accuracy and efficiency. Real-world tests further confirm its robustness and adaptability in handling irregular and noisy measurements. The flexible design of DispFormer makes it a versatile tool for various geophysical and engineering applications.
    
%TC:endignore

%% ------------------------------------------------------------------------ %%
%
%                               Introduction
%
%% ------------------------------------------------------------------------ %%
\section{Introduction}
    % the importance of surface tomography and it's application
    Surface wave tomography, which utilizes the cross-correlation functions of long-term seismic ambient noise \cite{shapiro_2005_Highresolution, yao_2006_Surfacewave, nishida_2009_Global, sager_2018_Full} or surface waves generated by regional and global earthquakes \cite{montagner_1990_Global, shapiro_2002_Montecarlo, yang_2006_Regional, li_2015_Love, babikoff_2019_LongPeriod, herrmann_2021_ShortPeriod}, has proven to be a powerful method for investigating the interior of the Earth. This technique is widely used for high-resolution imaging of the crust and lower mantle \cite{kaviani_2020_Crustal, xiao_2022_Simultaneous, liu_2023_Highresolution}, and is increasingly applied in near-surface exploration \cite{socco_2004_Surfacewave, mordret_2013_Nearsurface}. A commonly used approach in surface wave tomography involves a two-step inversion process. In the first step, group and/or phase velocity maps are constructed across multiple periods. These maps are then employed to generate dispersion curves for each grid cell, which are subsequently inverted to obtain one-dimensional (1-D) shear wave velocity ($v_s$) profiles \cite{barmin_2001_Fast}.

    % the traditional surface wave methodology and it's limitation 
    The inversion process that maps surface wave dispersion curves to a 1-D depth profile of $v_s$ is inherently nonlinear and underdetermined \cite{xia_1999_Estimation, dalmoro_2007_Joint}. Conventional methods include linearized inversion techniques \cite{herrmann_2013_Computer}, which iteratively refine an assumed initial velocity model using gradient descent, and global search algorithms such as Monte Carlo \cite{socco_2008_Improved, maraschini_2010_Monte} and particle swarm optimization (PSO) \cite{song_2012_Application}, which explore a broad parameter space to identify optimal solutions. As the number of deployed seismic arrays and shared databases increases, the limitations of both approaches become more pronounced. Linearized inversion encounters difficulties in yielding accurate results without a good initial model, while global search algorithms face significant computational challenges \cite{sambridge_2002_MONTE, sen_2013_Global}.

    % deep leaning based surface wave dispersion curve inversion methodology and it's limitation
    Deep learning has emerged as a promising alternative to traditional methods, offering a balance between efficiency and accuracy in various inversion applications, including gravity inversion \cite{li_2022_SelfSupervised, zhang_2022_Deep}, electromagnetic inversion \cite{colombo_2021_Physicsdriven, liu_2022_PhysicsDriven}, and seismic inversion \cite{yang_2019_Deeplearning, li_2020_DeepLearning, li_2025_Simultaneous, li_2025_Unsupervised}. In the context of surface wave dispersion curve inversion, early studies used fully connected neural networks (FCNNs) to estimate surface wave velocities and layer thicknesses \cite{meier_1993_Initial, devilee_1999_Efficient, meier_2007_Global}, paving the way for more advanced deep learning approaches. For example, \citeA{hu_2020_Using} utilized convolutional neural networks (CNNs) to improve inversion results on two regional datasets from continental China and southern California. \citeA{earp_2020_Probabilistic} and \citeA{yang_2022_Reconstruction} employed mixture density networks to derive $v_s$ structures while quantifying inversion uncertainty. \citeA{aleardi_2021_Hybrid}, along with \citeA{gan_2024_Deep}, used residual networks (ResNets) to directly map the full dispersion spectrum to $v_s$ models. Additionally, \citeA{luo_2022_Constructing} trained a deep FCNN on a global synthetic dataset and validated its performance on regional datasets. \citeA{cai_2022_SemiSupervised} proposed a semi-supervised Cycle-GAN to enhance generalization in poorly constrained regions. 

    % illuminate the challenge for traditional deep learning method and the complex of observed dispersion curve
    Despite these advancements, most existing network architectures are limited by the requirement for fixed-length dispersion data, which restricts their applicability to real-world scenarios where dispersion curves often face challenges such as inconsistent data ranges, missing values, and low signal-to-noise ratios \cite{socco_2004_Surfacewave, deng_2022_Extended, qin_2022_HighResolution}. Moreover, these models typically exhibit limited generalization capabilities, performing well on training datasets but underperforming on unseen or diverse datasets \cite{hu_2020_Using, gan_2024_Deep}. In practice, applying these models often requires large labeled datasets for case-specific training and the alignment of training and observed data through methods such as cropping, interpolation, and padding. However, large labeled datasets and length-aligned observations are rarely available in surface wave dispersion curve inversion studies. These challenges underscore the need for more robust and adaptable methods that can accommodate varying lengths and deliver reliable results across a broad range of datasets.
    
    % proposed methods for deal with vary-length and zero-shot/few-shot usage
    To address these challenges, this study introduces three primary contributions: 1) A network architecture supporting varying length data: DispFormer leverages a transformer-based structure to process dispersion data independently at each period, without requiring data alignment or fixed-length inputs. The model encodes dispersion data using linear layers and position embeddings, extracts period-related features through multiple transformer blocks, and finally projects the results into a 1-D velocity profile. 2) Incorporation of physical constraints during the training process: DispFormer integrates depth-sensitive physical constraints into the training process, ensuring that the network respects the underlying geophysical principles of surface wave dispersion. The pre-trained model, trained on a global synthetic dataset, handles varying-length data effectively and generalizes well to regional datasets. This enables zero-shot testing, allowing DispFormer to generate reliable initial models that can effortlessly integrate with traditional inversion workflows. 3) Pre-training for improved data efficiency and transferability: By pre-training on a global synthetic dataset, DispFormer incorporates prior knowledge of surface wave dispersion, reducing dependence on large labeled datasets. This pre-training enhances the model’s ability to perform well in few-shot scenarios, where only a small amount of labeled data is available. Both synthetic and real-world experiments demonstrate that fine-tuning the pre-trained model significantly improves inversion accuracy, often outperforming traditional global search methods.

%% ------------------------------------------------------------------------ 
%
%                              Methodology
%
%% ------------------------------------------------------------------------ 
\section{Methodology}
    
    \subsection{Surface Wave Dispersion Curve Inversion} 
        % the basis of surface wave forward modeling
        For a horizontally layered Earth model, the forward modeling of the Rayleigh wave dispersion curves can be expressed as \cite{thomson_1950_Transmission, haskell_1953_Dispersion}:
        \begin{equation}
            \mathbf{d}(\tau) = G(v_p, v_s, \rho, h),
        \end{equation}
        where $G$ is the forward operator that takes the layered earth model as input, including P-wave velocity ($v_p$), S-wave velocity ($v_s$), density ($\rho$), and layer thickness ($h$) for each layer. The output of this operation, $\mathbf{d}(\tau)$, corresponds to the dispersion data at a given period $\tau$, which includes both phase and group velocities.
    
        Inversion aims to estimate the underlying Earth model parameters based on observed dispersion data. Traditional inversion methods can be broadly categorized into linearized and global search approaches. Linearized inversion iteratively refines an initial model, typically using gradient-based optimization to minimize the discrepancy between observed and modeled data \cite{xia_1999_Estimation, herrmann_2013_Computer}. However, its effectiveness heavily depends on the quality of the initial model; a well-chosen initial model increases the likelihood of convergence to the global minimum, while a poor choice raises the risk of being trapped in a local minimum \cite{tarantola_1984_Linearized}. Global search algorithms, in contrast, explore the parameter space more comprehensively to identify the optimal model. These methods are less dependent on initial guesses and are generally more robust but come with substantial computational costs, which limits their practicality for large-scale applications. 
    
        % deep learning mappings between dispersion curve and 1-D velocity profiles
        To achieve a balance between efficiency and accuracy in inversion, deep learning techniques have been increasingly employed to learn the nonlinear mappings between dispersion curves and S-wave velocity \cite{chen_2024_Surface, gan_2024_Deep}, which can be mathematically expressed as:
        \begin{equation} 
            \mathbf{m} = f(\mathbf{d}; \theta) 
        \end{equation}
        where $f$ denotes the neural network, parameterized by $\theta$. The inputs to the neural network, $\mathbf{d}$, consists of phase and/or group velocity dispersion data. The outputs, $\mathbf{m} = [\mathbf{v}_p, \mathbf{v}_s, \mathbf{\rho}, \mathbf{h}]$, represents the predicted subsurface velocity model. In practice, empirical relationships are often used to estimate $v_p$ and $\rho$ due to their relatively low sensitivity to Rayleigh wave dispersion data \cite{xia_1999_Estimation}. To further streamline the inversion process, the model can be divided into thin layers of equal thickness, a strategy validated in previous studies \cite{hu_2020_Using, cai_2022_SemiSupervised, wang_2023_Deeplearningbased}. Within the period range and model depth considered in this study, research shows that thin layers (0.1, 0.25, 0.5 km) have minimal impact on inversion accuracy, whereas excessively thick layers ($\geq$ 2 km) significantly degrade accuracy \cite{hu_2020_Using}. These simplifications allow the inversion process to focus on $v_s$, the primary variable of interest for surface wave tomography. 
        
        Supervised learning techniques are then used to train the neural network by minimizing the discrepancy between the predicted and true velocity models. This process is achieved through the following misfit functions:
        \begin{equation} 
        \mathcal{J}(\mathbf{d}, \mathbf{m}; \theta) = \frac{1}{N} \sum_{i=1}^{N} \mathcal{L}(f(d_i; \theta), \mathbf{m}_i) 
        \end{equation}
        where $\mathcal{J}$ denotes the misfit function, $N$ represents the number of samples, $f(\mathbf{d}_i; \theta)$ is the predicted velocity model for the $i$-th input dispersion data $\mathbf{d}_i$, and $\mathbf{m}_i$ is the corresponding true velocity model. The function $\mathcal{L}(\cdot, \cdot)$ quantifies the difference between the predicted and true velocity models, typically using metrics such as mean squared error (MSE) or mean absolute error (MAE). Once trained, the neural network can efficiently conduct inversion mapping without requiring iterative optimization or extensive random sampling, which are common in traditional methods.

    \begin{figure}[!ht]
        \centering
        \includegraphics[width=0.70\textwidth]{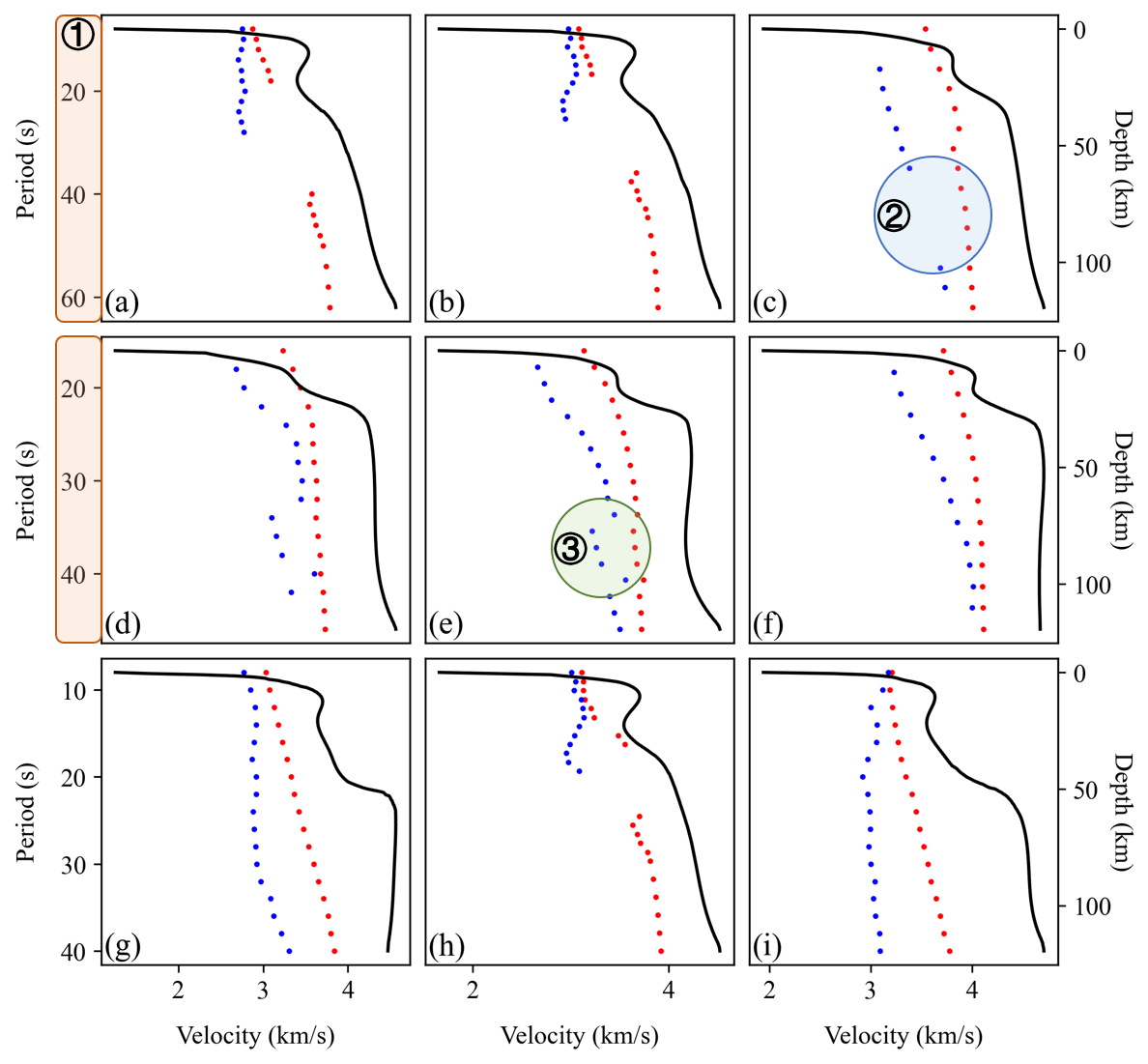}
        \caption{Examples of observed dispersion curves and the corresponding S-wave velocity profile derived from the China Seismological Reference Model \cite{xiao_2024_CSRM10}. The black line indicate the 1-D S-wave velocity profiles, with a thickness of 0.5 km, spanning from 0 to 120 km. The blue and red scatters indicate the observed phase and group dispersion data. Remarkable features of the real dispersion curves are specifically labeled in the figure, including (1) varying period ranges, (2) missing values, and (3) lower signal-to-noise ratios.}
        \label{fig1}
    \end{figure}
    
    \subsection{DispFormer: Inversion of Dispersion Curves with Variable Lengths}
        
        Previous studies have successfully applied various neural networks, such as FNNs, CNNs, and ResNets, to learn the inversion mapping between dispersion curves and S-wave velocity profile. However, the intrinsic characteristics of the dispersion data pose challenges for training a universal neural network that can adapt to diverse real-world scenarios. A critical factor is the varying period ranges, which correlate with the inversion depth, as longer periods typically reveal deeper structures \cite{socco_2004_Surfacewave,socco_2010_Surfacewave}. These period ranges can differ widely across study areas, and even within a single dataset, the dispersion curves may have distinct effective ranges. Additionally, some dispersion curves suffer from low signal-to-noise ratios or missing values due to limitations in the data acquisition and processing. As shown in Figure~\ref{fig1}, observed dispersion curves often display such variations, complicating the task of training a neural network that can generalize effectively across diverse scenarios. Consequently, neural network architectures that rely on fixed-length input data may struggle to handle these inconsistencies.
    
\begin{figure}[!ht]
            \centering
            \includegraphics[width=\textwidth]{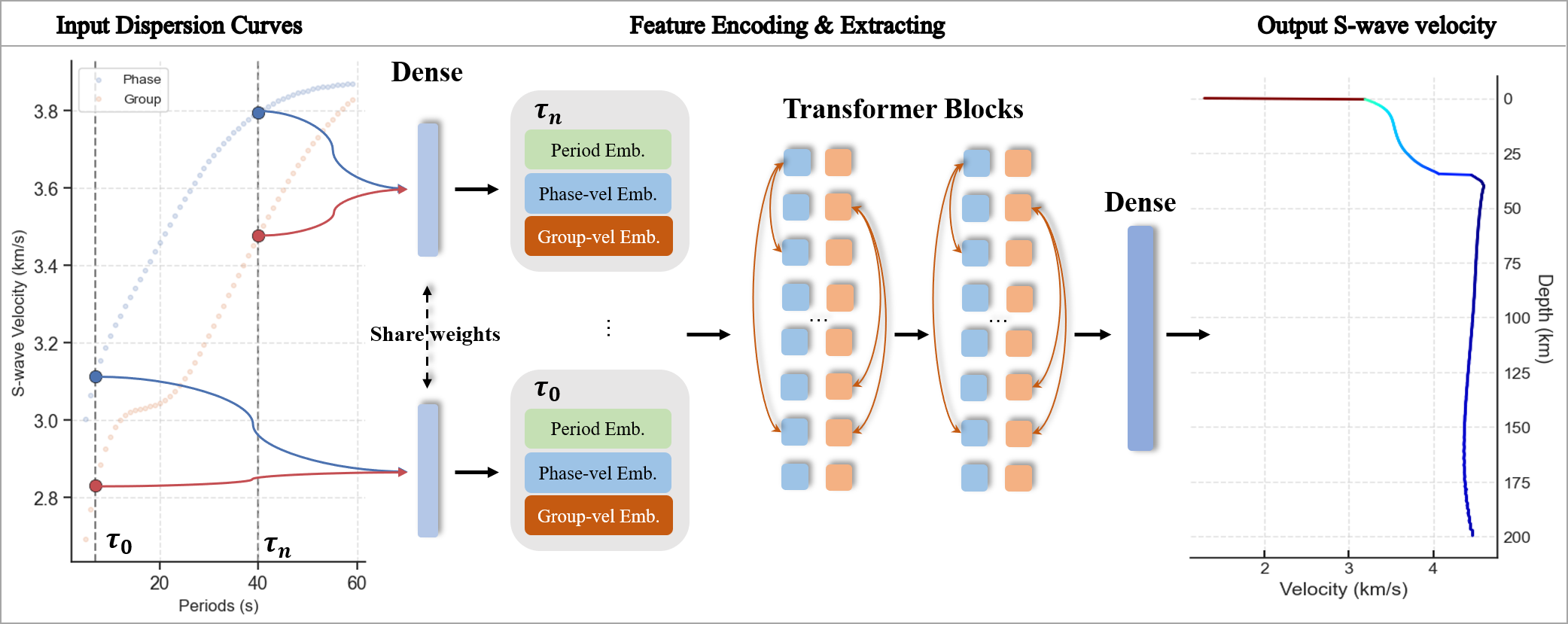}
            \caption{Architecture of the DispFormer network. The network takes dispersion curves, including period, phase velocity, and group velocity, as inputs. Each period of the dispersion data is encoded using linear layers combined with positional embeddings to retain temporal relationships. Transformer blocks are then employed to model inter-period correlations. Finally, the extracted features are projected through a dense layer to estimate the shear-wave velocity ($v_s$) at fixed depths.}
            \label{fig2}
        \end{figure}

        In this study, we propose DispFormer, a transformer-based architecture designed to handle dispersion data of arbitrary length. Figure~\ref{fig2} illustrates the workflow of DispFormer, where dispersion curves serve as input, and the corresponding S-wave velocity profile at each depth is generated as output. Initially, dense layers separately encode the period, phase velocity, and group velocity for each period. To preserve the relative positions between periods, position embeddings based on the period information are incorporated. Transformer blocks are then employed to capture the relationships between different periods of the surface wave dispersion curves, which are essential for modeling the depth-dependent velocity structure. Finally, a dense layer maps the extracted features to the S-wave velocity. The flexibility of DispFormer in handling data of arbitrary length is achieved through two strategies: encoding each period of data independently and leveraging the transformer architecture. A more detailed structure of the DispFormer is shown in Supporting Figure S1.
        
    \subsection{DispFormer Pre-training and Fine-tuning Workflow}
        
\begin{figure}[!ht]
            \centering
            \includegraphics[width=0.85\textwidth]{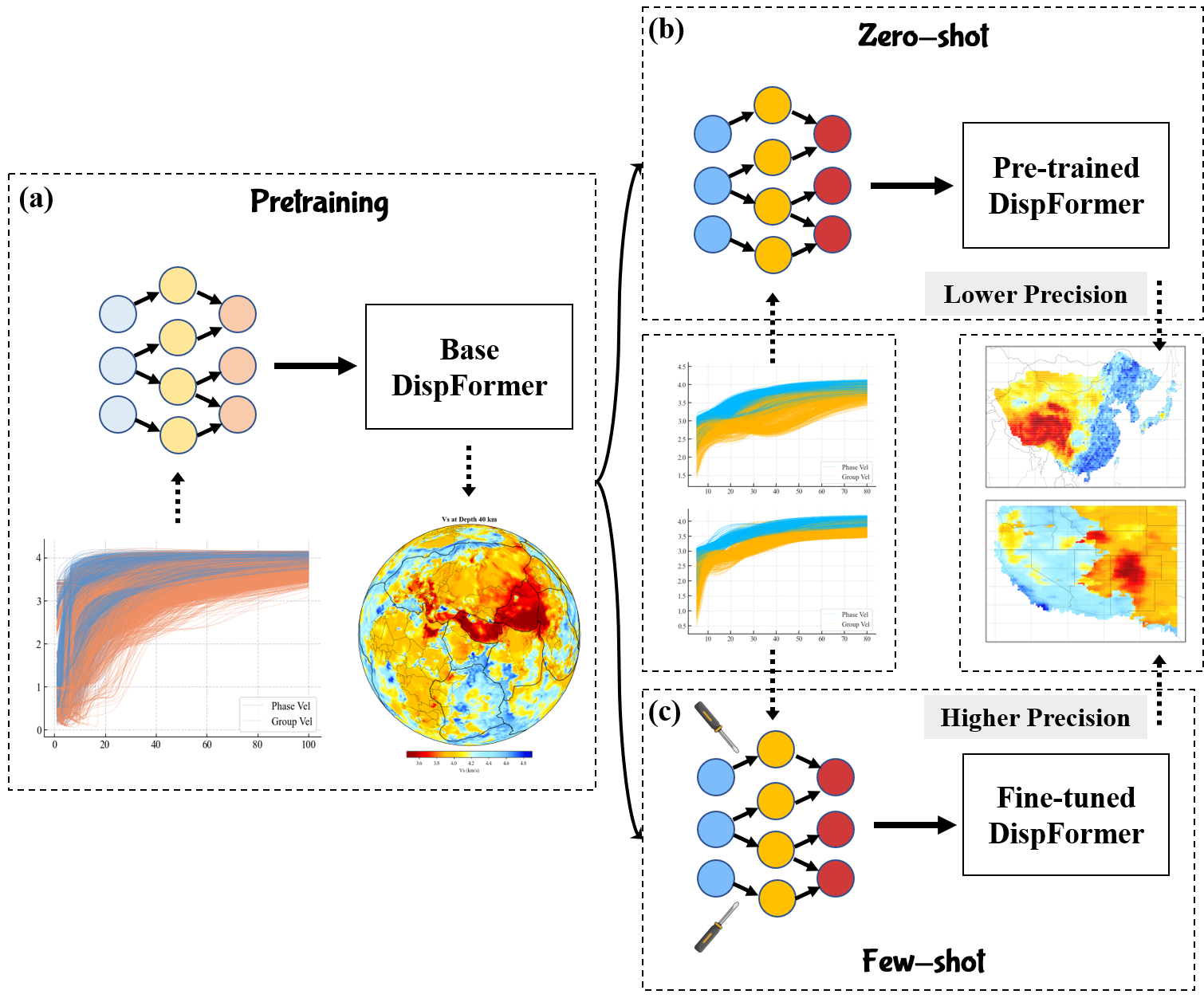}
            \caption{Workflow of DispFormer pre-training and utilization strategies. (a) Pre-training process using a global synthetic dataset to train the model for general inversion tasks. (b) Zero-shot inversion using the pre-trained model applied directly to regional datasets without labeled data. (c) Fine-tuning process, where a small amount of labeled data is used to adapt the pre-trained model, improving accuracy on specific regional datasets.}
            \label{fig3}
        \end{figure}

        The capability of DispFormer to accommodate dispersion data of arbitrary length facilitates the implementation of a pre-training and fine-tuning strategy, significantly enhances its generalization ability. In this study, a global synthetic dataset was created for pre-training. This dataset, with spatial resolution of approximately 1 $\degree$ and period ranges spanning from 1 to 100 seconds, is designed to capture crust and upper mantle structures down to depths of about 200 km (Figure~\ref{fig3}a). The resulting pre-trained model provides a robust foundation for subsequent applications.
    
        In regional studies with higher spatial resolutions (e.g., $<$ 0.25 $\degree$) and varying period ranges (e.g., 10-60 s, 8-80 s), the pre-trained DispFormer can be directly applied to map observed dispersion curves to S-wave velocity, even in the absence of labeled data. This "zero-shot" strategy (Figure~\ref{fig3}b) eliminates the need for region-specific training, offering an efficient solution for generating velocity profiles. While the inversion results may not always outperform those of traditional global search methods due to the domain gap between pre-training dataset and regional observation datasets, the velocity models generated by the pre-trained DispFormer remain highly valuable. They can serve as practical initial models or provide meaningful constraints for subsequent inversion processes. In this way, DispFormer functions as a plug-and-play tool to support and enhance traditional inversion techniques.
    
        When labeled data, such as well logs or inversion results from global search methods, is available, the pre-trained model can be fine-tuned to better align with the regional data distribution. This "few-shot" strategy (Figure~\ref{fig3}c) requires only a small subset of labeled data, yet it significantly enhances inversion accuracy. Remarkably, fine-tuned DispFormer achieves inversion results that are comparable to, or even surpass, those of traditional global search methods.

    \subsection{DispFormer Model Training and Performance Assessment}
        
        During the training process, DispFormer iteratively updates model parameters, $\theta$, by minimizes the misfit function, $\mathcal{J}$. Since the velocity varies with depth, the misfit function is defined as the normalized mean squared error (NMSE) to account for differences in relative velocity scales. This is expressed as \cite{chen_2022_Deep}:
        \begin{equation}
            \mathcal{L}(f(\mathbf{d}; \theta), \mathbf{m}) = \frac{1}{N} \sum_{i=1}^{N} (\frac{f(\mathbf{d}_i; \theta) - \mathbf{m}_i}{\mathbf{m}_i})^2
        \end{equation}
        where $N$ is the number of training samples, and $\mathbf{m}_i$ and $f(\mathbf{d}; \theta)$ denote the ground truth and predicted velocity profiles for the $i$-th sample, respectively. The NMSE ensures that training is less sensitive to variations in absolute velocity values across different depths.

        To optimize the model, the adaptive moment estimation (Adam) optimizer \cite{kingma_2017_Adam} is employed, balancing fast convergence and stable updates. The initial learning rate is set to $1 \times 10^{-4}$ during pre-training and reduced to $5 \times 10^{-5}$ for fine-tuning. A step-based learning rate decay scheduler (StepLR) progressively decreases the learning rate as training advances, mitigating overfitting and improving convergence stability. 
    
        Model performance is evaluated using the mean absolute error, which provides an intuitive measure of prediction accuracy by quantifying the absolute deviation between predicted and true velocity models:
        \begin{equation} 
            MAE = \frac{1}{N} \sum_{i=1}^{N} | f(\mathbf{d}_i; \theta) - \mathbf{m}_i |. 
        \end{equation}
        While NMSE serves as the primary training objective, MAE offers a complementary metric for assessing predictive accuracy.

%% ------------------------------------------------------------------------ 
%
%                         PROCEDURE FOR DATA GENERATION
%
%% ------------------------------------------------------------------------ 

\section{Synthetic and Real-World Datasets}

    \subsection{Global Synthetic Datasets for Model Pre-training}

        The pre-training datasets used in this study are entirely synthetic, constructed from an extensive collection of 1-D velocity profiles. To capture diverse and realistic features, a paired global velocity-dispersion dataset is first created based on the updated Earth crust and lithosphere model (LITHO1.0) \cite{pasyanos_2014_LITHO10}. The fundamental mode Rayleigh-wave phase and group velocity are computed from these extracted velocity profiles, and the construction process is outlined as follows:
        \begin{enumerate}
            
            \item Initially, 40,962 1-D S-wave velocity profiles, extending to a depth of 200 km, are extracted from the LITHO1.0 database. Any water layers in the profiles are removed, and each profile is then converted into an isothermal layer model with a uniform layer thickness of 0.5 km, using linear interpolation.
    
            \item Given the relatively low sensitivity of Rayleigh wave phase and group velocities to $v_p$ and $\rho$, $v_p$ for depths above 120 km is computed from $v_s$ using the empirical relationships established by Brocher \cite{brocher_2005_Empirical}, and with a fixed $v_p/v_s$ ratio of 1.79 for depths between 120 and 200 km \cite{kennett_1995_Constraints}. Additionally, $\rho$ is derived from $v_p$ using Brocher's empirical relationship \cite{brocher_2005_Empirical}.
    
            \item Theoretical Rayleigh wave phase and group velocity dispersion curves for periods ranging from 1 to 100 seconds are generated using the Computer Programs in Seismology (CPS) software package \cite{herrmann_2013_Computer}. To ensure comprehensive coverage of real-world scenarios, periods are drawn not only uniformly at integer period positions but also logarithmically and randomly within the period domain \cite{wang_2023_Deeplearningbased}.
    
            \item During the pre-training phase, the complete dataset is used to train the model, and the best optimized model is selected as the base model. To improve the generalization and stability of the base model, several data augmentation strategies are introduced, including: a) adding random Gaussian noise (approximately 5\%), b) zeroing out random segments of the data (approximately 10\%), and c) randomly removing either phase velocity or group velocity.
    
            \item Considering that dispersion curves with varying period ranges are sensitive to different depth ranges, a dynamic calculation approach is adopted to determine the approximate depth range based on the wavelengths of observed dispersion curves. The calculation can be formulated as:
            \begin{equation}\label{eq6}
                z_{min} = min (C_1 \lambda_{p},\; C_2 \lambda_{g}),
            \end{equation}
            \begin{equation}\label{eq7}
                z_{max} = C_3 \; max (\lambda_{p},\; \lambda_{g}),
            \end{equation}
            where $z_{min}$ and $z_{max}$ represent the minimum and maximum sensitive depths, respectively. The phase and group wavelengths for each period $\tau_i$ are given by $\lambda_p^i = v_{phase}^i \times \tau_i$ and $\lambda_g^i = v_{group}^i \times \tau_i$, where $v_{phase}^i$ and $v_{group}^i$ are the phase and group velocities. The constants $C_1$, $C_2$, and $C_3$ are empirical scaling factors, with values of $C_1 = 1/3$, $C_2 = 1/2$, and $C_3 = 1.1$ used in this study.
            
        \end{enumerate}
        
        DispFormer is pre-trained on the complete LITHO1.0 dataset for 5000 iterations, and the model with the lowest training loss is selected for subsequent zero-shot and few-shot testing. Detailed training loss profiles are presented in Figure S2.
    
    \subsection{Regional Synthetic Datasets for Model Validation}
    
        To evaluate the stability and performance of DispFormer, two regional synthetic datasets are generated based on the S-wave velocity profiles from \citeA{shen_2013_3D} and \citeA{shen_2016_Seismic}. The dataset from \citeA{shen_2013_3D} consists of 6,803 1-D S-wave velocity profiles derived from a tomographic model of the central and western United States, while \citeA{shen_2016_Seismic} provides 4,527 profiles from continental China. These datasets are herein referred to as the Central and Western US Dataset (CWD) and the Continental China Dataset (CCD), respectively. The thickness, P-wave velocity and density for both datasets are calculated using the same parametrization strategy as that employed for the pre-training dataset. The dispersion periods are sampled at 1-second intervals, spanning 10 to 60 s for the CWD and 5 to 80 s for the CCD, providing distinct period ranges compared to the pre-training data.
        
\begin{figure}[!ht]
            \centering
            \includegraphics[width=0.65\textwidth]{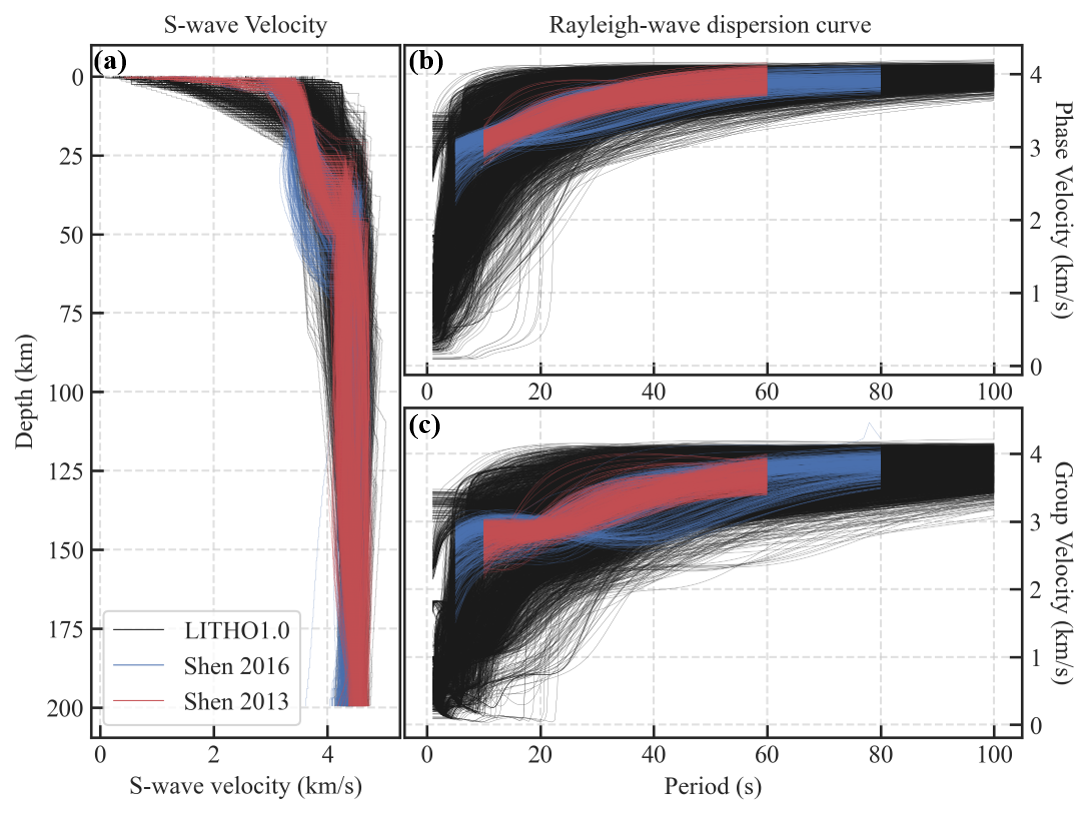}
            \caption{Distribution of the global and local synthetic datasets. (a) S-wave velocity profiles, with the black, red, and blue curves representing data from LITHO1.0, \citeA{shen_2013_3D}, and \citeA{shen_2016_Seismic}, respectively; (b) distribution of Rayleigh-wave phase velocity dispersion curves; (c) distribution of Rayleigh-wave group velocity dispersion curves.}
            \label{fig4}
        \end{figure}

        When testing DispFormer on these synthetic datasets, the zero-shot strategy allows direct evaluation without requiring any training data. In contrast, the few-shot strategy involves selecting a small subset (less than 2\%) of the regional dataset for fine-tuning, done through a hierarchical selection process. The distribution of global and regional synthetic datasets is illustrated in Figure~\ref{fig4}, and the training and validation loss curves for few-shot learning on the CWD and CCD datasets are shown in Figure S3.

    \subsection{Field Datasets for Model Testing}
        
        \citeA{xiao_2024_CSRM10} propose a high-resolution China Seismological Reference Model (CSRM) by integrating various seismic data, with extensive surface wave dispersion data being used to derive the S-wave velocity. The dispersion data, spanning periods from 8 to 70 seconds, are extracted from three-component waveforms of 9,361 teleseismic events recorded at 4,193 seismic stations across mainland China. These observations are mapped onto a regular grid, with finer resolution of 0.2 \degree $\times$ 0.2 \degree for the North-South Seismic Belt and the trans-North China orogen regions, and coarser resolution of 0.4 \degree $\times$ 0.4 \degree for the remainder of the continent.
        
        For this study, 12,705 observed dispersion curves are extracted from the original CSRM database and directly used to construct the test dataset. Figure~\ref{fig1} presents examples from the observed dataset, highlighting issues such as varying period ranges, missing values, and low signal-to-noise ratios. Additionally, the corresponding reference velocity profiles were interpolated into 0.5 km thick layers, with a maximum depth of 120 km. For the few-shot learning test on the CSRM dataset, 38 and 114 samples are selected for fine-tuning, with the corresponding training loss curves shown in Figure S4. Notably, to avoid introducing errors inherent in the inversion results of the reference model into network training, the paired fine-tuning dataset consists of the reference velocity model and its corresponding synthetic dispersion curves generated by the CPS program, rather than real dispersion curves and the reference model. For more general applications, the reference velocity models can be obtained from well log data or generated using global search algorithms, followed by constructing paired fine-tuning datasets according to the aforementioned procedure.

%% ------------------------------------------------------------------------ 
%
%                        APPLICATION AND VERIFICATION
%
%% ------------------------------------------------------------------------ 

\section{Application and Verification}

    \subsection{Zero-shot Generation of Reasonable Initial Models}
    
        This section accesses the zero-shot performance of the pre-trained DispFormer on two regional synthetic datasets. Given that DispFormer is designed to process dispersion data of arbitrary length, it directly take in dispersion curves from the CWD and CCD datasets, with period ranges of 10 to 60 s for CWD and 5 to 80 s for CCD, to generate the corresponding inverted S-wave velocity profiles. 
        
\begin{figure}[!ht]
            \centering
            \includegraphics[width=0.75\textwidth]{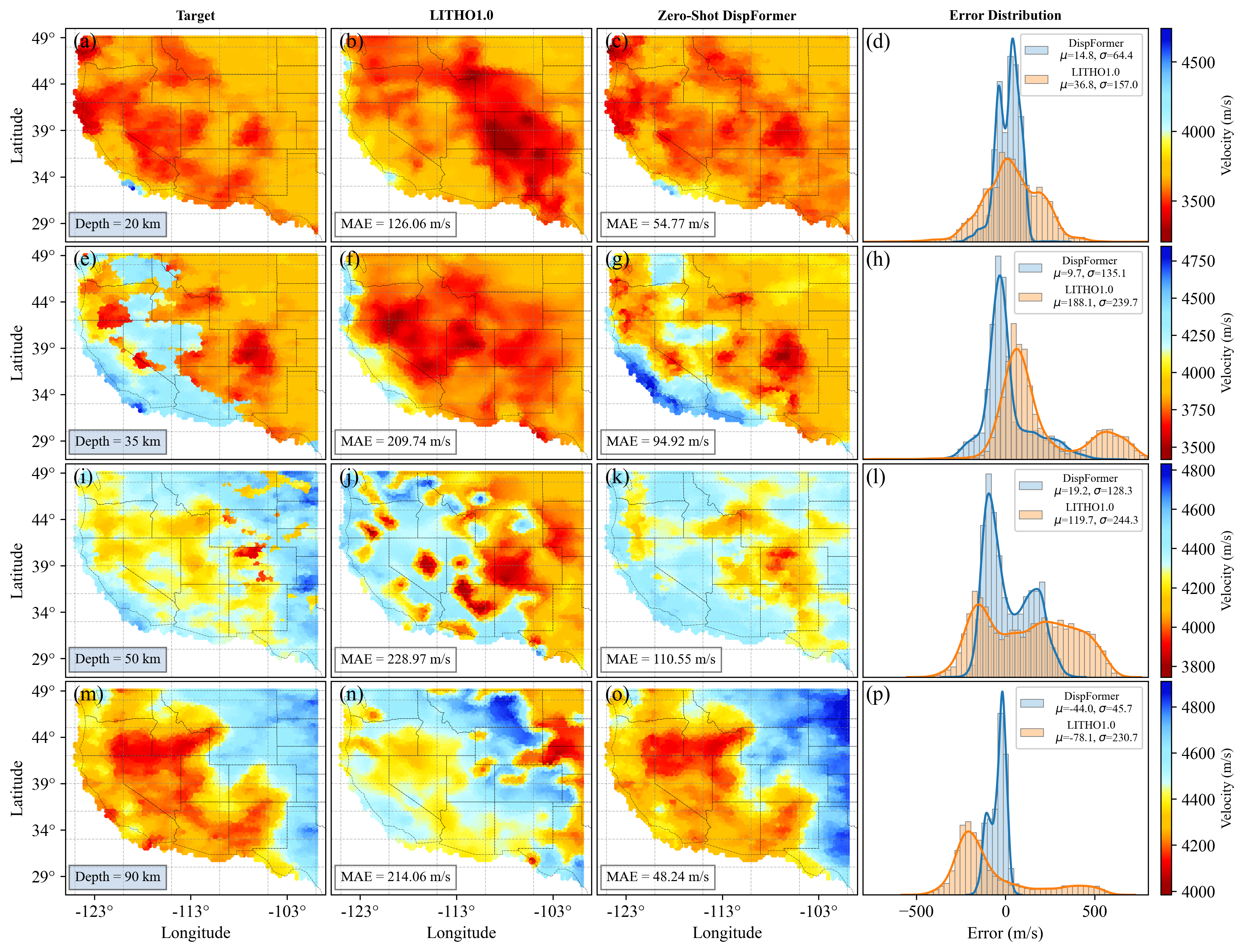}
            \caption{Comparison of S-wave velocity models at depths of 20, 40, 50, and 90 km using the CWD dataset. Columns show (1) target velocity model, (2) interpolated LITHO1.0 model, (3) inverted result from zero-shot DispFormer, and (4) error distributions relative to the target model.}
            \label{fig5}
        \end{figure}

\begin{figure}[!ht]
            \centering
            \includegraphics[width=0.85\textwidth]{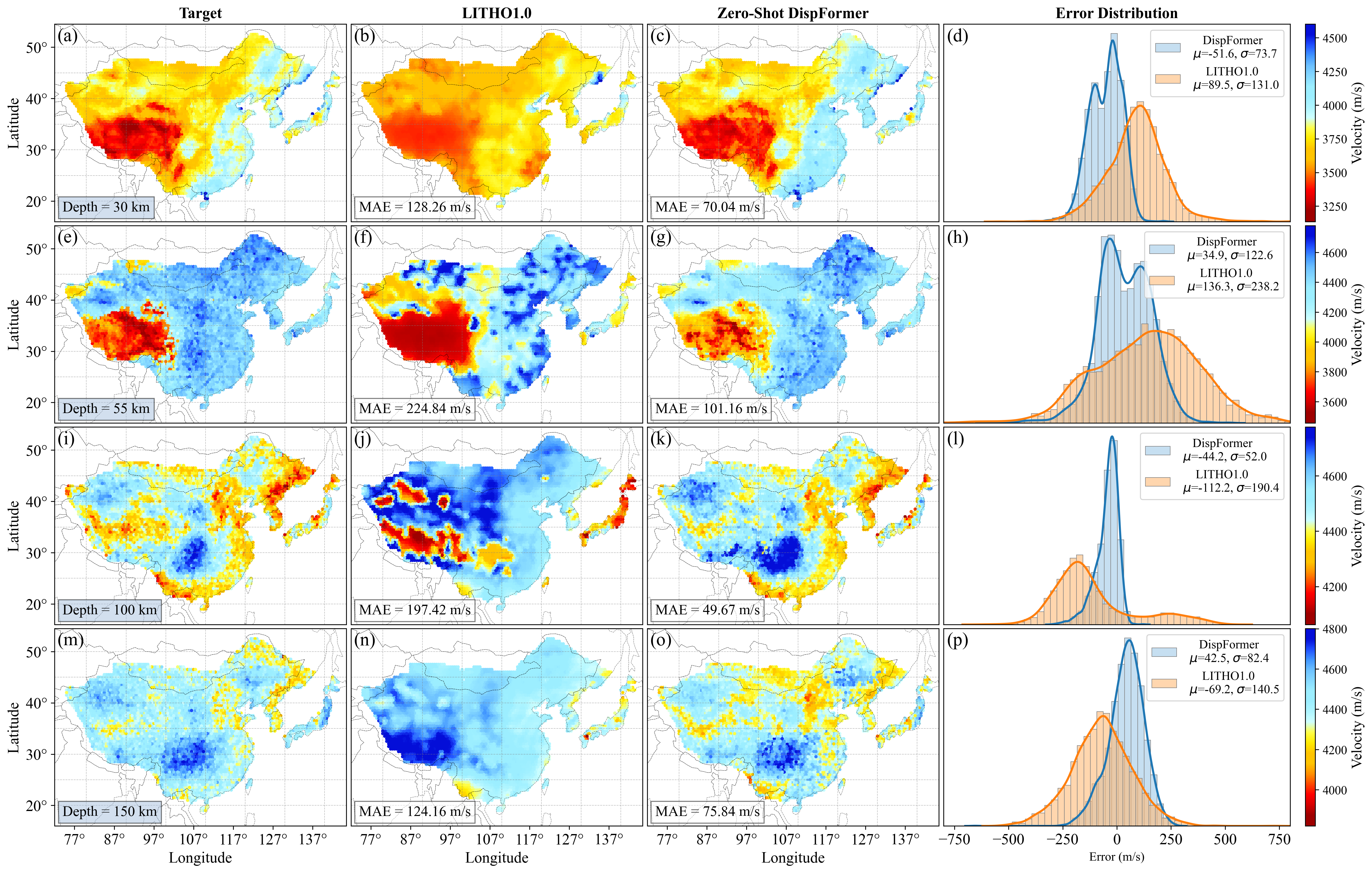}
            \caption{Comparison of S-wave velocity models at depths of 30, 55, 100 and 170 km using the CCD dataset. Columns show (1) target velocity model, (2) interpolated LITHO1.0 model, (3) inverted result from zero-shot DispFormer, and (4) error distributions relative to the target model.}
            \label{fig6}
        \end{figure}

        Figure~\ref{fig5} shows slices of the true and inverted S-wave velocity models at depths of 20, 40, 50, and 90 km from the CWD dataset. The first column (Figs. ~\ref{fig5}a, e, i, and m) shows the ground-truth velocity model, while the second column shows the interpolated model derived from the LITHO1.0 database, which is also used for pre-train the DispFormer model. The third column presents the inversion results obtained using the pre-trained DispFormer in a zero-shot manner, and the fourth column visualizes the error distributions between the target model and both the interpolated LITHO1.0 model and the zero-shot DispFormer’s results. It is important to clarify that the comparison with the interpolated LITHO1.0 model is not intended to assess the capability of DispFormer in replicating its training data. Rather, the major objective is to evaluate the effectiveness of the zero-shot DispFormer in providing a plausible initial model, given that LITHO1.0-based interpolated models are conventionally employed as initial guesses in linear iterative or global search inversion schemes. As shown in the fourth column, the error distribution of the zero-shot DispFormer is more tightly clustered around zero, with fewer large deviations compared to the interpolated model. For the CCD dataset, which has a broader period range, the model achieves deeper S-wave velocity inversions. Figure~\ref{fig6} shows the true and inverted velocity models at depths of 30, 55, 100, and 170 km. The zero-shot DispFormer effectively captures low-velocity anomalies in the Tibetan Plateau region, underscoring its capability to provide accurate structural details at greater depths.
    
        Table~\ref{table1} quantifies the MAE between the target model and both the interpolated LITHO1.0 model and the zero-shot DispFormer results for the CWD and CCD datasets. The zero-shot DispFormer consistently outperforms the interpolated model across all depths, achieving significantly lower MAE values. This further supports the proposed approach of using the zero-shot DispFormer to generate initial models, which has the potential to enhance the accuracy and efficiency of subsequent inversion processes.

        \begin{table}[!ht]
            \caption{MAE of the interpolated LITHO1.0 model and the zero-shot DispFormer on the CWD and CCD datasets.}
            \centering
            \begin{tabular}{c|c|c}
            \hline
            \multicolumn{1}{c|}{}       & Interped LITHO1.0 & zero-shot DispFormer \\ \hline
            \multicolumn{1}{l|}{CWD}    & 234.53 m/s        & \textbf{72.78 m/s}   \\
            \multicolumn{1}{l|}{CCD}    & 233.61 m/s        & \textbf{73.27 m/s}   \\ \hline
            \end{tabular}
            \label{table1}
        \end{table}
    
    \subsection{Few-shot DispFormer for Accurate Inversion}
    
        The adaptability of a pre-trained DispFormer model to local datasets can be further enhanced through fine-tuning when labeled data is accessible. To evaluate the effectiveness of few-shot learning, experiments are carried out on two regional datasets. The inversion results obtained by fine-tuning with limited labeled data are contrasted with those from a global search algorithm. For the global search baseline, a particle swarm optimization (PSO) algorithm is implemented \cite{luu_2023_Evodcinv}, with a search range of ± 0.6 km/s around the true velocity model and an iteration limit of 2000.
    
\begin{figure}[!ht]
            \centering
            \includegraphics[width=0.75\textwidth]{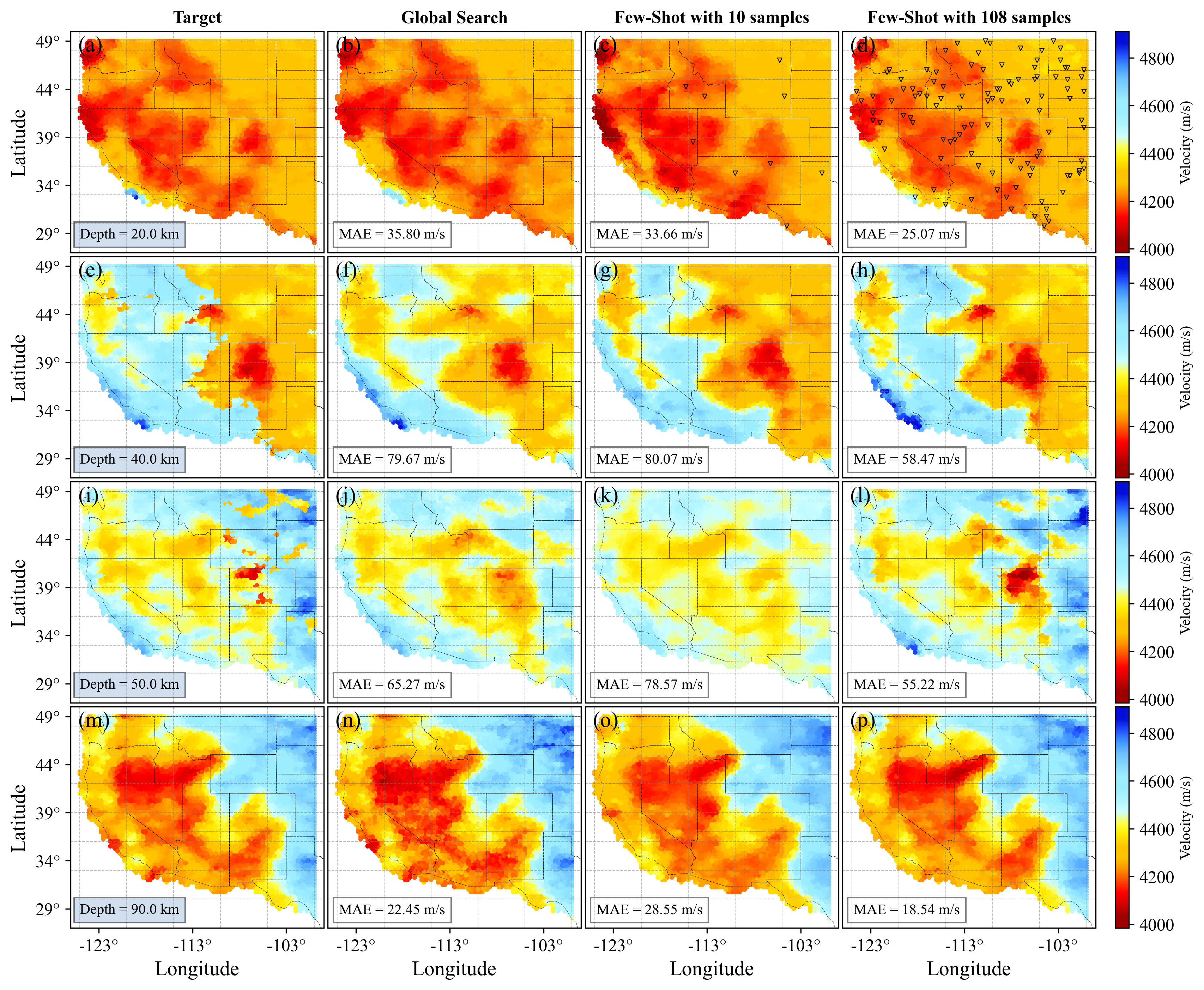}
            \caption{Comparison of inverted S-wave velocity models at depths of 20, 40, 50, and 90 km using the CWD dataset. Columns show: (1) target velocity model, (2) PSO global search results, (3) fine-tuned DispFormer with 10 samples, and (4) fine-tuned DispFormer with 108 samples.}
            \label{fig7}
        \end{figure}

\begin{figure}[!ht]
            \centering
            \includegraphics[width=0.85\textwidth]{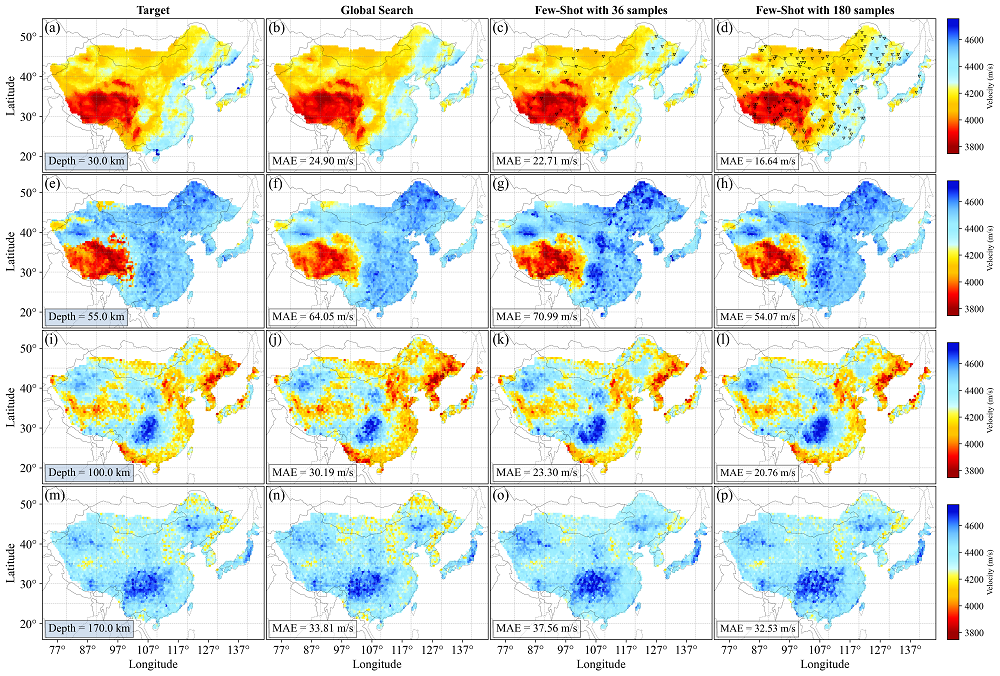}
            \caption{Comparison of inverted S-wave velocity models at depths of 30, 55, 100, and 170 km using the CCD dataset. Columns show: (1) target velocity model, (2) PSO global search results, (3) fine-tuned DispFormer with 36 samples, and (4) fine-tuned DispFormer with 180 samples.}
            \label{fig8}
        \end{figure}

        Figure~\ref{fig7} compares the inverted S-wave velocity models obtained from the PSO and few-shot DispFormer at depths of 20, 40, 50, and 90 km using the CWD dataset. The first column shows the true S-wave velocity model, while the second column presents the inversion results using the PSO method. The third and fourth columns display the inverted results from the few-shot DispFormer, fine-tuned with 10 and 108 labeled samples, respectively. Similarly, Figure~\ref{fig8} provides a comparison using the CCD dataset, with fine-tuning performed using either 36 or 180 labeled samples. Table~\ref{table2} summarizes the MAE between the target velocity model and the inversion results obtained from the PSO global search method and the few-shot DispFormer. These results demonstrate that the few-shot DispFormer, which uses only a small number of labeled samples (e.g. 10 for CWD and 36 for CCD) for fine-tuning, achieves inversion results comparable to those from the global search method. Furthermore, when hundreds of labeled samples are available, the fine-tuned DispFormer outperforms the global search method. These findings highlight the potential of few-shot DispFormer to efficiently generate high-quality inversion models, making it particularly useful for practical applications where labeled data is scarce.

        To comprehensively evaluate computational efficiency, we measured the training, fine-tuning, and inference times of DispFormer using an NVIDIA A100-PCIE-40GB GPU and a 128-core CPU server (HUAWEI, Kunpeng 920). Pretraining on the global dataset (LITHO1.0) with 40,962 samples for 5000 iterations required approximately 37 hours. Fine-tuning on the CWD dataset using 10 and 108 labeled samples for 500 iterations took 210 and 224 seconds, respectively. In contrast, the PSO global search method, processing the entire CWD dataset with 6803 samples, required approximately 159 hours. Although the PSO implementation did not utilize multi-CPU parallelization, which could potentially reduce runtime, its computational cost would still likely exceed the pretraining duration of DispFormer. For inference, the fine-tuned DispFormer model processed the entire dataset in approximately 1 second, underscoring its significant computational advantage over the global search method.

        \clearpage
        
        \begin{table}[!ht]
            \caption{MAE of the PSO and the few-shot DispFormer models on the CWD and CCD datasets. The numbers in parentheses indicate the number of samples used for fine-tuning.}
            \centering
            \begin{tabular}{c|c|c|c}
            \hline
            \multicolumn{1}{c|}{}   & PSO       & \begin{tabular}[c]{@{}c@{}}Few-shot DispFormer\\ (10 CWD / 36 CCD)\end{tabular} & \begin{tabular}[c]{@{}c@{}}Few-shot DispFormer\\ (108 CWD / 180 CCD)\end{tabular} \\ \hline
            \multicolumn{1}{l|}{CWD} & 42.77 m/s & 45.70 m/s                    & \textbf{33.84 m/s}            \\
            \multicolumn{1}{l|}{CCD} & 36.87 m/s & 36.58 m/s                    & \textbf{28.69 m/s}            \\ \hline
            \end{tabular}
            \label{table2}
        \end{table}
    
    \subsection{Real-world Application: DispFormer on CSRM Dataset}
    
\begin{figure}[!ht]
            \centering
            \includegraphics[width=1.0\textwidth]{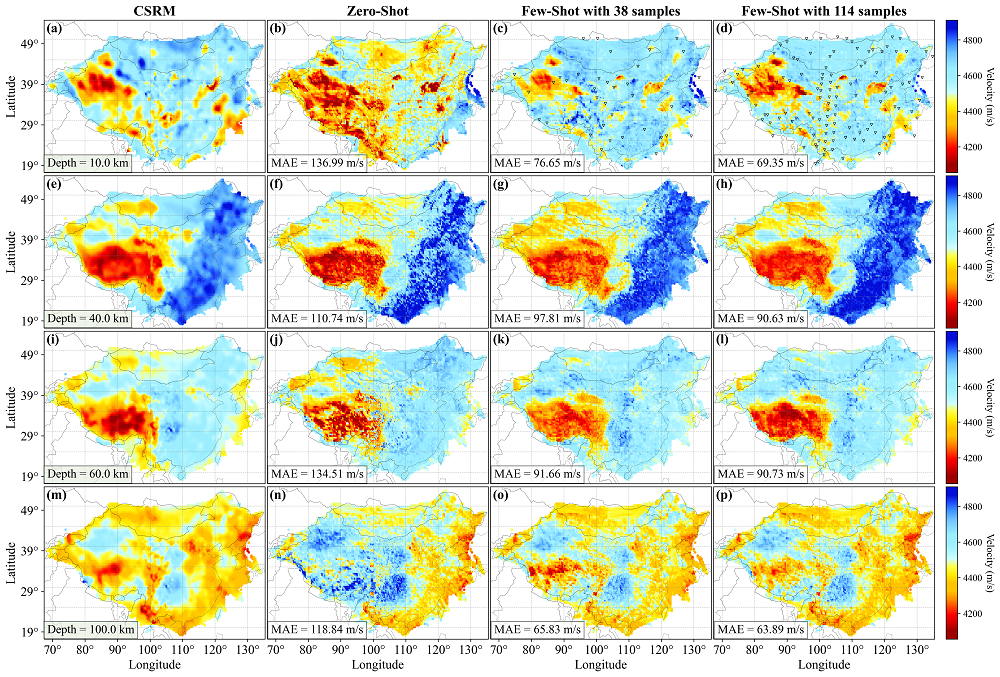}
            \caption{Comparison of the reference model and inverted results using zero-shot and few-shot DispFormer at depths of 10, 45, 60, and 100 km on the CSRM dataset. Columns show: (1) reference model from CSRM, (2) zero-shot DispFormer, (3) fine-tuned DispFormer with 38 samples, and (4) fine-tuned DispFormer with 114 samples.}
            \label{fig9}
        \end{figure}

        To evaluate the performance of DispFormer in practical scenarios, we tested the pre-trained model via both zero-shot and few-shot approaches on the CSRM dataset. Figure~\ref{fig9} presents the reference model alongside the inversion results at depths of 10, 40, 60, and 100 km. The first column displays the reference model from \citeA{xiao_2024_CSRM10}, followed by the results obtained using zero-shot DispFormer in the second column. The third and fourth columns illustrate the inverted results after fine-tuning with 38 and 114 labeled samples, respectively. Notably, the labeled samples employed for fine-tuning were randomly selected from the reference model, and their corresponding dispersion curves were synthesized using the CPS program. The MAE between the inversion results and the reference model is reported in the lower-left corner of each sub-figure. The results indicate that zero-shot DispFormer effectively reconstructs large-scale structural features, such as the prominent low-velocity zone beneath the Tibetan Plateau. Furthermore, fine-tuning with as few as 1\% labeled data can significantly improve accuracy, particularly in resolving finer structural details.
    
\begin{figure}[!ht]
            \centering
            \includegraphics[width=1.0\textwidth]{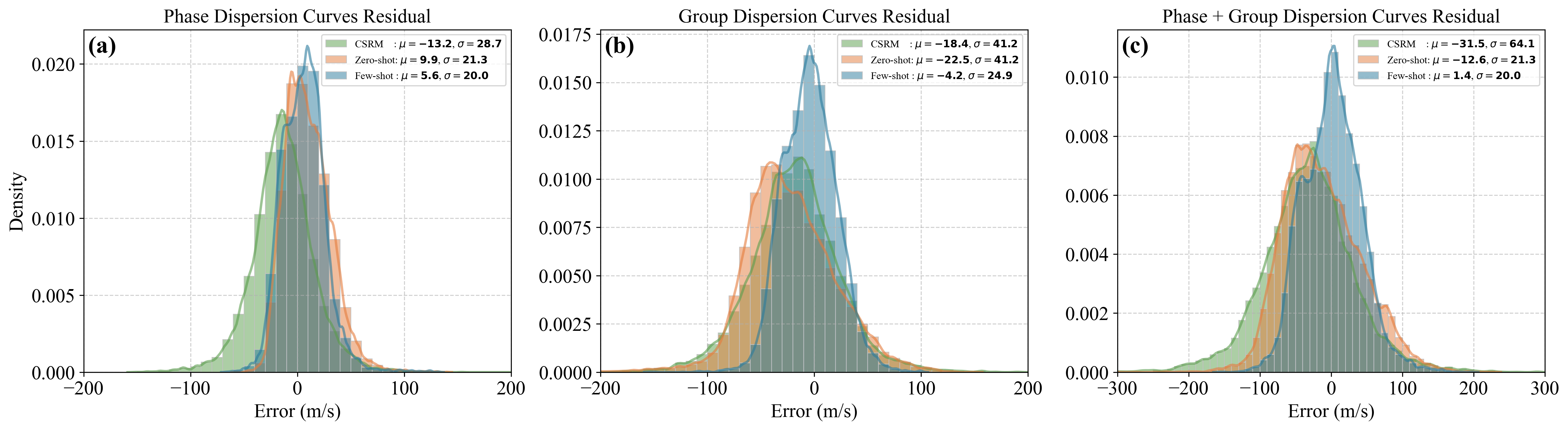}
            \caption{Comparison of data residuals between observed dispersion curves and those synthesized from the reference model (green), zero-shot DispFormer (pink), and few-shot DispFormer (blue) inversion results. Panel (a) shows the phase velocity error distribution, panel (b) presents the group velocity error distribution, and panel (c) illustrates the combined error distribution for both phase and group velocities.}
            \label{fig10}
        \end{figure}

        In real-world applications, where no ground-truth model is available for direct validation, inversion accuracy is commonly assessed through data residuals (i.e., the differences between synthetic dispersion curves derived from the inversion results and the actual observed dispersion curves). To this end, Figure~\ref{fig10} compares the data residual distributions for the CSRM reference model, zero-shot DispFormer, and few-shot DispFormer fine-tuned with 114 labeled samples. The comparison reveals that the inversion results from zero-shot DispFormer closely resemble the CSRM reference model in the data domain, while few-shot DispFormer further reduces residuals, achieving a closer fit to the observed data. These findings underscore the capability of DispFormer to effectively handle complex real-world datasets, demonstrating its potential as a robust and adaptable tool for large-scale geophysical inversions across diverse applications.
    
\clearpage
%% ------------------------------------------------------------------------ 
%
%                        DISSCUSSION
%
%% ------------------------------------------------------------------------ 

\section{Discussion}

    \subsection{Dynamic Depth Alignment for Physically Consistent Inversion}
    
        The depth alignment strategy employed during training dynamically aligns the period ranges of the dispersion curves with their corresponding sensitivity depths, thereby significantly improving the zero-shot performance of DispFormer. This strategy relies on empirical formulas for $z_{min}$ (Equation~\ref{eq6}) and $z_{max}$ (Equation~\ref{eq7}) to compute the inversion depth range, allowing the model to capture depth-specific features that align with the physical sensitivity of the observed dispersion data. In traditional deep learning methods, fixed inversion depths are commonly used. However, these may not effectively align with the specific depths that are most sensitive to the observed data, potentially leading to inaccurate inversion results, particularly when applied to regional datasets with varying subsurface conditions. In contrast, by dynamically calculating the range of inversion depths, this strategy narrows the solution space and enables the model to learn more physically relevant features. 
        
        As shown in Table~\ref{table3}, the MAE comparison between zero-shot DispFormer models trained with and without depth alignment reveals that the alignment strategy consistently reduces MAE across two regional synthetic datasets. By aligning the learning process with the underlying geophysical principles of wave physics, this approach ensures that the model effectively captures depth-sensitive features, improving its performance in practical applications. These improvements emphasize the importance of incorporating physical constraints into the training process, which not only enhances the generalization ability of the neural network but also improves its explainability.
        
        The effectiveness of the dynamic depth alignment strategy as a soft physical constraint is governed by its hyperparameters $C_1$, $C_2$ and $C_3$. Narrower sensitivity ranges may provide insufficient constraints for deeper or shallower regions, while broader ranges may degrade generalization by incorporating extraneous depth features. In this study, the hyperparameters adopted were systematically calibrated to balance these trade-offs; however, further adjustments may be necessary for broader applications, such as shallow subsurface exploration or lower mantle inversion. Additionally, recent studies have proposed integrating sensitivity kernels to impose more stringent physical constraints during training \cite{chen_2024_Surface}. Although such approaches could enhance depth sensitivity alignment, they require kernel computation for each dispersion point, leading to high computational cost, especially for complex or multi-layered models. Therefore, addressing the integration of dynamic sensitivity kernels within the training framework remains a compelling avenue for future investigation.
    
        \begin{table}[!ht]
            \caption{Comparison of MAE for zero-shot DispFormer with and without depth alignment on the CWD and CCD datasets. The bolded values indicating the lower MAE achieved using the depth-aligned version.}
            \centering
            \renewcommand{\arraystretch}{1.0}
            \begin{tabular}{c|c|c}
            \hline
            \multicolumn{1}{c|}{}       & zero-shot (w/o Depth-Align) & zero-shot (w/ Depth-Align) \\ \hline
            \multicolumn{1}{l|}{CWD}    & 98.53 m/s                   &  \textbf{72.78} m/s            \\
            \multicolumn{1}{l|}{CCD}    & 91.20 m/s                   &  \textbf{73.27} m/s            \\ \hline
            \end{tabular}
            \label{table3}
        \end{table}

    \subsection{DispFormer Robustness to Noise and Missing Data}
        
        The previous synthetic tests demonstrated that, under ideal conditions with clean data, few-shot DispFormer can be fine-tuned with a limited dataset to achieve inversion results surpassing traditional methods. However, real-world data is often contaminated by noise, which can degrade the performance and stability of inversion models. To evaluate DispFormer’s robustness against noise, Gaussian white noise with a standard deviation ranging from 0\% to 10\% of the observed dispersion curves were added to two regional synthetic datasets. Table~\ref{table4} shows the MAE of the inversion results obtained from the fine-tuned DispFormer across varying noise levels. While the MAE increases with higher noise levels, the overall increase remains moderate, with errors staying below 20 m/s. These results highlight that DispFormer can still deliver accurate and stable inversion outcomes, even when noise is present in the data.

        In addition to noise robustness, real-world datasets often contain missing data due to instrument limitations or manual picking errors, which pose significant challenges to traditional deep learning frameworks. To evaluate DispFormer's robustness against missing data, this study simulated missing points in the observed dispersion curves by randomly removing up to 70\% of the data in both regional synthetic datasets. The resulting inversion performance, quantified by MAE, is presented in Table~\ref{table5}. Even with a substantial proportion of missing data (e.g., 50\%), DispFormer exhibited strong resilience, with minimal performance degradation compared to the complete datasets. These findings indicate that DispFormer effectively handles incomplete data, highlighting its applicability to real-world scenarios where data gaps are prevalent. Supporting Figures S5 to S10 provide visualizations of the inverted results under varying levels of Gaussian noise and missing data, further illustrating DispFormer's stability and accuracy.
    
        \begin{table}[!ht]
            \centering
            \renewcommand{\arraystretch}{1.0}
            \caption{Comparison of MAE for few-shot DispFormer under varying levels of Gaussian noise on the CWD and CCD datasets. The bolded values indicate the MAE under 0\% noise, representing the baseline performance for each configuration.}
            \begin{tabular}{c|l|l|l|l|l}
            \hline
            \diagbox[width=15em]{Few-Shot Samples}{Noise Level} & 0\%  & 2\%  & 5\%  & 8\%  & 10\%  \\
            \hline
            CWD with 10 samples  & \textbf{45.7} & 46.0 & 47.6 & 50.6 & 53.1  \\
            CWD with 108 samples & \textbf{33.8} & 35.1 & 40.2 & 47.4 & 52.6  \\
            CCD with 36 samples  & \textbf{36.5} & 37.2 & 40.3 & 45.5 & 49.9  \\
            CCD with 180 samples & \textbf{28.7} & 29.7 & 34.0 & 40.9 & 46.5  \\
            \hline
            \end{tabular}
            \label{table4}
        \end{table}

        \begin{table}[!ht]
            \centering
            \renewcommand{\arraystretch}{1.0}
            \caption{Comparison of MAE for few-shot DispFormer under varying levels of missing data on the CWD and CCD datasets. The bolded values indicate the MAE under 0\% missing, representing the baseline performance for each configuration.}
            \begin{tabular}{c|l|l|l|l|l}
            \hline
            \diagbox[width=15em]{Few-Shot Samples}{Missing level} & 0\%  & 10\%  & 30\%  & 50\%  & 70\%  \\
            \hline
            CWD with 10 samples  & \textbf{45.7} & 50.0 & 51.2 & 57.9 & 77.0  \\
            CWD with 108 samples & \textbf{33.8} & 35.7 & 39.1 & 47.2 & 72.0  \\
            CCD with 36 samples  & \textbf{36.5} & 45.8 & 46.6 & 48.9 & 69.3  \\
            CCD with 180 samples & \textbf{28.7} & 38.2 & 40.7 & 41.4 & 65.5  \\
            \hline
            \end{tabular}
            \label{table5}
        \end{table}

    \subsection{Uncertainty evaluation for the inverted velocity model}
    
        Observed dispersion curves often contain significant errors due to environmental noise during data acquisition and the subjectivity involved in manual picking. These errors can propagate into the inversion results, affecting their reliability. Therefore, assessing the impact of data uncertainty on inversion outcomes is a crucial aspect of inversion research \cite{lai_2005_Propagation, griffiths_2016_Surfacewave}. Bayesian inversion is a common method for uncertainty estimation, incorporating uncertainty directly into the inversion process \cite{gouveia_1998_Bayesian, shen_2013_Joint}, but it tends to be computationally expensive. Alternatively, Monte Carlo-based approaches estimate uncertainty by sampling and performing multiple inversions, transferring data errors to the model space \cite{shapiro_2002_Montecarlo, socco_2008_Improved, liu_2024_Multimodal}.
    
\begin{figure}[!ht] 
            \centering 
            \includegraphics[width=1.0\textwidth]
            {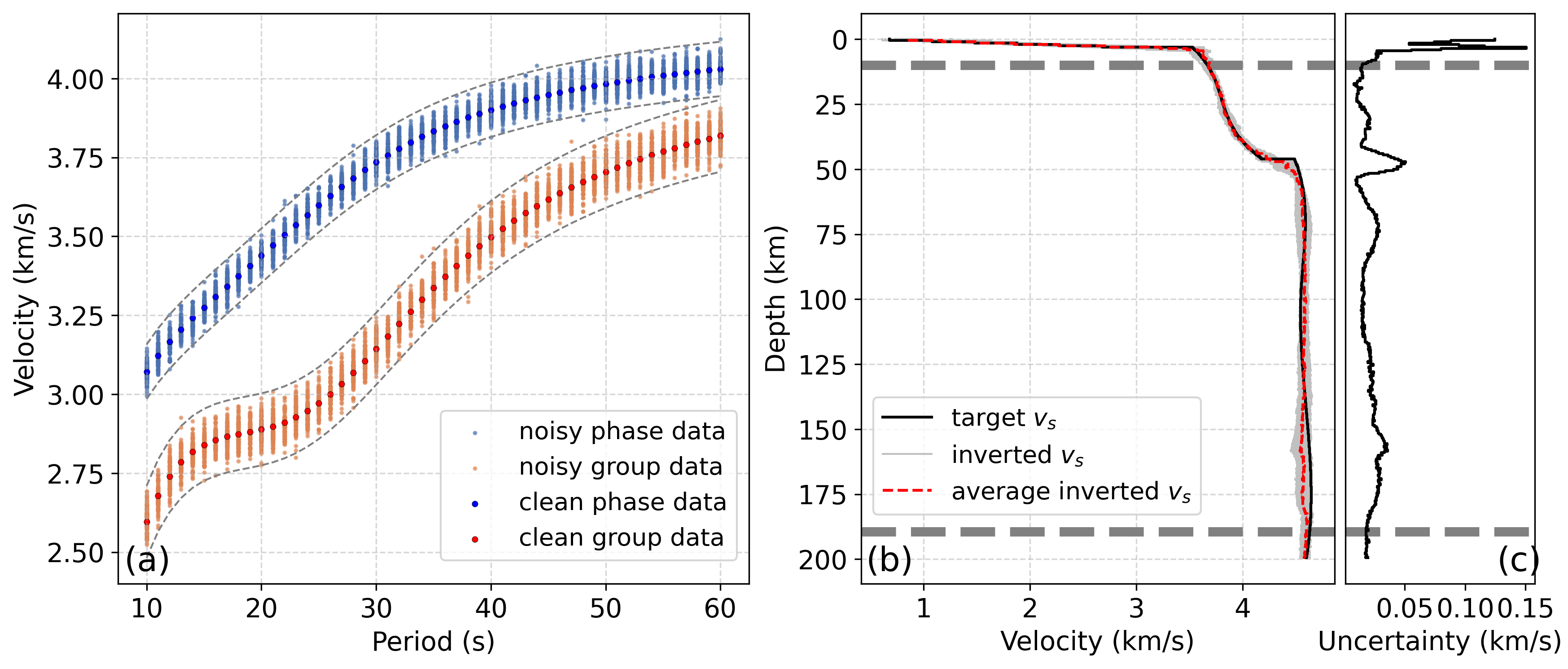} \caption{Uncertainty estimation by perturbing observed data at a single station from the CWD datasets. (a) Clean and noisy dispersion curves, with noisy data generated by adding Gaussian white noise at 10 \% standard deviation to the observed curves, repeated 1000 times. (b) Monte Carlo-based uncertainty estimation, the silver line shows the inverted $v_s$ using the fine-tuned DispFormer, the red line represents the averaged inverted results, and the black line denotes the ground truth. (c) Estimated uncertainty (standard deviation) for each layer.} 
            \label{fig11} 
        \end{figure}

        However, Monte Carlo-based methods require many sampling and inversions, leading to high computational costs. In contrast, end-to-end deep learning methods like DispFormer are much more efficient, enabling thousands of mappings in seconds, making them ideal for uncertainty assessment. To demonstrate this, we applied DispFormer to assess the uncertainty in inversion results for a single station from the CWD dataset. The few-shot DispFormer was fine-tuned with 108 samples, and the test dataset was generated by independently adding random Gaussian white noise (10\% of the standard deviation of the entire observed dispersion curve) to each period of the observed data. This process was repeated 1,000 times, with the noise being re-generated for each repetition. The clean and noisy data are shown in Figure~\ref{fig11}a, the inversion results are presented in Figure~\ref{fig11}b, and the estimated uncertainty is illustrated in Figure~\ref{fig11}c. Despite significant errors in the observed data, the inversion results from DispFormer remain concentrated in a narrow range, with the average inversion results closely matching the ground-truth model. Notably, larger errors are observed at shallower depths (0–7 km), primarily due to the low sensitivity of the dispersion data to velocities at these depths. The most reliable inversion range, determined by the proposed depth alignment strategy, is marked by a short-thick dotted line in Figs. \ref{fig11}b and \ref{fig11}c.
    
    \subsection{DispFormer Generalization, Limitations, and Potential Extensions}
    
\begin{figure}[!ht]
            \centering
            \includegraphics[width=0.80\textwidth]{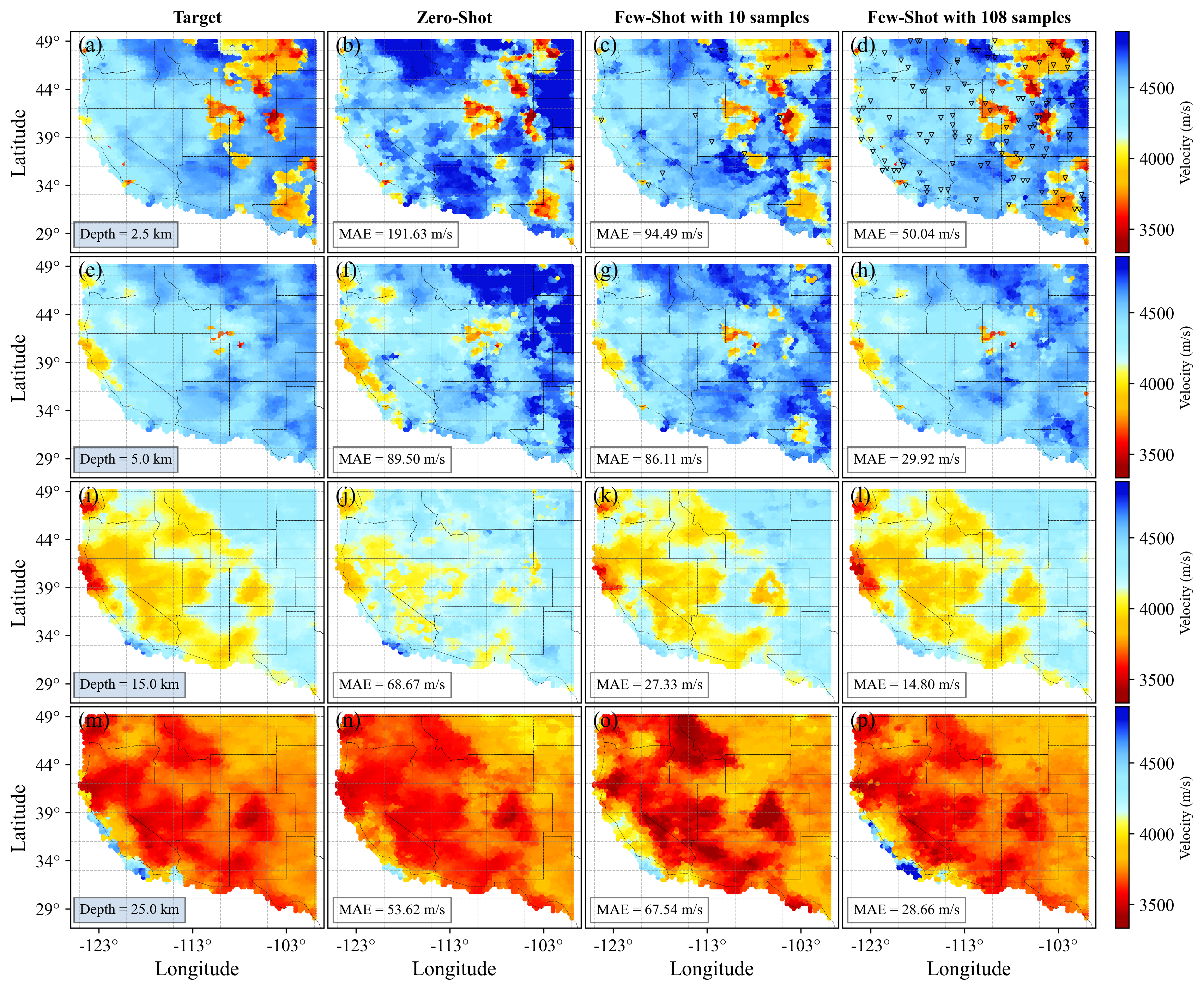}
            \caption{Comparison of the inversion results using zero-shot and few-shot DispFormer at depths of 2.5, 5, 15, and 25 km on a new CWD dataset with a period range of 1–10 s. The columns from left to right show: (1) target velocity model, (2) zero-shot DispFormer, (3) few-shot DispFormer fine-tuned with 10 samples, and (4) few-shot DispFormer fine-tuned with 108 samples.}
            \label{fig12}
        \end{figure}

        In the application and validation sections, the DispFormer model pre-trained on the LITHO1.0 dataset was applied to datasets with varying period ranges, including CWD (10–60 s), CCD (5–80 s), and CSRM (8–70 s), targeting large-scale structures in the lower crust and mantle. The model exhibited robust transferability across these datasets. However, ambient noise studies using short-period nodal arrays typically focus on dispersion periods below 10 s, presenting a more challenging scenario for the pre-trained model. To assess its generalization to shallow subsurface structures, a new CWD dataset with periods of 1–10 s was constructed, primarily probing depths within 30 km. Zero-shot and few-shot tests were conducted using the same methodology as in previous sections. Figure~\ref{fig12} presents horizontal slices of the S-wave velocity model at depths of 2.5 km, 5 km, 15 km, and 25 km, showing the true model, zero-shot results, and few-shot results with 10 and 108 samples, respectively. The results indicate that despite the shift to shorter periods, the model maintains stable inversion performance, particularly after few-shot fine-tuning.

        Despite these promising results, two key challenges limit its broader applicability: (1) Resolution Limitations: The uniform 0.5 km layer thickness constrains the model’s ability to resolve fine-scale structures, particularly in engineering applications and near-surface inversions. (2) Geological Complexity: In shallow crust or urban exploration, some geophysical structures exhibit extremely complex velocity contrasts that are not adequately represented in the LITHO1.0 dataset. Addressing these limitations provides a pathway for further enhancing the DispFormer model. Reconstructing the pre-training dataset with finer or variable layer thicknesses would improve inversion resolution. Additionally, the targeted collection of more comprehensive datasets with diverse geological settings would enable broader applications to complex geological scenarios. Extending the model to support multi-physics joint inversion, such as integrating surface wave and receiver function data, could further enhance inversion accuracy. Beyond these adjustments, DispFormer’s capacity to integrate heterogeneous data can be leveraged for broader geophysical applications, including magnetotelluric, gravity, magnetic, and receiver function inversions. These advancements would position DispFormer as a versatile framework for geophysical inversion in diverse geological settings.

%% ------------------------------------------------------------------------ 
%
%                        CONCLUSION
%
%% ------------------------------------------------------------------------ 
\section{Conclusion}

    This study introduces DispFormer, a transformer-based neural network designed to invert $v_s$ models from Rayleigh-wave phase and/or group dispersion curves. DispFormer processes dispersion data independently for each period, leveraging transformer blocks to extract period-specific features and subsequently mapping them to $v_s$ profiles. The architecture is specially designed to handle dispersion data of arbitrary length, making it directly applicable to real-world datasets with varying lengths, without requiring modifications to the network structure or alignment of training and test data. Additionally, a dynamic depth alignment strategy is employed during training, incorporating physical constraints based on the sensitivity of dispersion data to different depths. This ensures that the model effectively captures depth-sensitive features, thereby enhancing its performance in practical applications.

    To evaluate its effectiveness, DispFormer was pre-trained on the global synthetic LITHO1.0 dataset and tested on two regional synthetic datasets in both zero-shot and few-shot modes. The synthetic tests demonstrate that even without labeled data, zero-shot DispFormer generates inversion profiles that outperform the interpolated reference model used as the pretraining target, making it a reliable tool for producing high-quality initial models. When fine-tuned with partial labeled data, the few-shot mode surpasses traditional global search methods in inversion accuracy, emphasizing DispFormer's efficiency in generating reliable models with constrained data. Real case studies further confirm DispFormer's versatility and generalizability, as the model adapts well to complex datasets with varying period ranges, missing values, and low signal-to-noise ratios. Moreover, DispFormer demonstrates strong robustness to noise and missing data, maintaining stable and accurate inversion results even when a high level of Gaussian noise is added or a substantial portion of the observed data is removed. Additionally, its efficient uncertainty quantification, enabled by the end-to-end deep learning framework, significantly reduces computational costs and time compared to traditional methods such as Monte Carlo or Bayesian inversion.
    
    In summary, DispFormer offers a powerful solution for surface wave tomography. Its zero-shot capability for generating improved initial models, few-shot fine-tuning for superior inversion results, robustness to noisy data, and efficient uncertainty quantification make it a promising tool for a wide range of geophysical inversion applications.

%% ------------------------------------------------------------------------ %%
%
%                         Data Availability Statement
%
%% ------------------------------------------------------------------------ %%
%TC:ignore
\section*{Open Research Section}
    The open-source package DispFormer developed in this study is available at \citeA{liu_2025_Surface} and can also be accessed via GitHub at \url{https://github.com/liufeng2317/DispFormer}. All synthetic tests conducted in this study can be reproduced using the scripts available in the examples folder. We thank \citeA{pasyanos_2014_LITHO10} for making the updated Earth crust and lithosphere model (LITHO1.0) publicly available, \citeA{shen_2013_3D} and \citeA{shen_2016_Seismic} for their open-source velocity models for the central and western United States and continental China regions, and \citeA{xiao_2024_CSRM10} for releasing velocity models and real-world datasets for the continental China region. Using these publicly available datasets, we have curated and generated additional datasets that can be accessed at \citeA{liu_2025_Surface}.

%% ------------------------------------------------------------------------ %%
%
%                           Acknowledgments
%
%% ------------------------------------------------------------------------ %%
\section*{Acknowledgments}
    We sincerely thank editor Dr. Geoffrey Fox, the associate editor, and three anonymous reviewers for their valuable time and for providing constructive and insightful comments that have significantly improved this study. This work was supported by the Shanghai Artificial Intelligence Laboratory. We also acknowledge the Shanghai Artificial Intelligence Laboratory for providing computational resources and the Science Discovery Platform (Intern-Discovery; available at \url{https://discovery.intern-ai.org.cn/}) for offering the testing and demonstration environment.

%% ------------------------------------------------------------------------ %%
%
%                           References and Citations
%
%% ------------------------------------------------------------------------ %%
\bibliography{reference}

\begin{thebibliography}{}

\bibitem [\protect \citeauthoryear {%
Aleardi%
\ \BBA {} Stucchi%
}{%
Aleardi%
\ \BBA {} Stucchi%
}{%
{\protect \APACyear {2021}}%
}]{%
aleardi_2021_Hybrid}
\APACinsertmetastar {%
aleardi_2021_Hybrid}%
\begin{APACrefauthors}%
Aleardi, M.%
\BCBT {}\ \BBA {} Stucchi, E.%
\end{APACrefauthors}%
\unskip\
\newblock
\APACrefYearMonthDay{2021}{}{}.
\newblock
{\BBOQ}\APACrefatitle {A Hybrid Residual Neural Network--Monte Carlo Approach to Invert Surface Wave Dispersion Data} {A hybrid residual neural network--monte carlo approach to invert surface wave dispersion data}.{\BBCQ}
\newblock
\APACjournalVolNumPages{Near Surface Geophysics}{19}{4}{397--414}.
\newblock
\begin{APACrefDOI} \doi{10.1002/nsg.12163} \end{APACrefDOI}
\PrintBackRefs{\CurrentBib}

\bibitem [\protect \citeauthoryear {%
Babikoff%
\ \BBA {} Dalton%
}{%
Babikoff%
\ \BBA {} Dalton%
}{%
{\protect \APACyear {2019}}%
}]{%
babikoff_2019_LongPeriod}
\APACinsertmetastar {%
babikoff_2019_LongPeriod}%
\begin{APACrefauthors}%
Babikoff, J\BPBI C.%
\BCBT {}\ \BBA {} Dalton, C\BPBI A.%
\end{APACrefauthors}%
\unskip\
\newblock
\APACrefYearMonthDay{2019}{}{}.
\newblock
{\BBOQ}\APACrefatitle {Long-Period Rayleigh Wave Phase Velocity Tomography Using USArray} {Long-period rayleigh wave phase velocity tomography using usarray}.{\BBCQ}
\newblock
\APACjournalVolNumPages{Geochemistry, Geophysics, Geosystems}{20}{4}{1990--2006}.
\newblock
\begin{APACrefDOI} \doi{10.1029/2018GC008073} \end{APACrefDOI}
\PrintBackRefs{\CurrentBib}

\bibitem [\protect \citeauthoryear {%
Barmin%
, Ritzwoller%
\BCBL {}\ \BBA {} Levshin%
}{%
Barmin%
\ \protect \BOthers {.}}{%
{\protect \APACyear {2001}}%
}]{%
barmin_2001_Fast}
\APACinsertmetastar {%
barmin_2001_Fast}%
\begin{APACrefauthors}%
Barmin, M\BPBI P.%
, Ritzwoller, M\BPBI H.%
\BCBL {}\ \BBA {} Levshin, A\BPBI L.%
\end{APACrefauthors}%
\unskip\
\newblock
\APACrefYearMonthDay{2001}{}{}.
\newblock
{\BBOQ}\APACrefatitle {A Fast and Reliable Method for Surface Wave Tomography:} {A fast and reliable method for surface wave tomography:}.{\BBCQ}
\newblock
\APACjournalVolNumPages{Pure and Applied Geophysics}{158}{8}{1351--1375}.
\newblock
\begin{APACrefDOI} \doi{10.1007/PL00001225} \end{APACrefDOI}
\PrintBackRefs{\CurrentBib}

\bibitem [\protect \citeauthoryear {%
Brocher%
}{%
Brocher%
}{%
{\protect \APACyear {2005}}%
}]{%
brocher_2005_Empirical}
\APACinsertmetastar {%
brocher_2005_Empirical}%
\begin{APACrefauthors}%
Brocher, T\BPBI M.%
\end{APACrefauthors}%
\unskip\
\newblock
\APACrefYearMonthDay{2005}{}{}.
\newblock
{\BBOQ}\APACrefatitle {Empirical Relations between Elastic Wavespeeds and Density in the Earth's Crust} {Empirical relations between elastic wavespeeds and density in the earth's crust}.{\BBCQ}
\newblock
\APACjournalVolNumPages{Bulletin of the Seismological Society of America}{95}{6}{2081--2092}.
\newblock
\begin{APACrefDOI} \doi{10.1785/0120050077} \end{APACrefDOI}
\PrintBackRefs{\CurrentBib}

\bibitem [\protect \citeauthoryear {%
Cai%
, Qiu%
\BCBL {}\ \BBA {} Niu%
}{%
Cai%
\ \protect \BOthers {.}}{%
{\protect \APACyear {2022}}%
}]{%
cai_2022_SemiSupervised}
\APACinsertmetastar {%
cai_2022_SemiSupervised}%
\begin{APACrefauthors}%
Cai, A.%
, Qiu, H.%
\BCBL {}\ \BBA {} Niu, F.%
\end{APACrefauthors}%
\unskip\
\newblock
\APACrefYearMonthDay{2022}{}{}.
\newblock
{\BBOQ}\APACrefatitle {Semi-Supervised Surface Wave Tomography With Wasserstein Cycle-Consistent GAN: Method and Application to Southern California Plate Boundary Region} {Semi-supervised surface wave tomography with wasserstein cycle-consistent gan: Method and application to southern california plate boundary region}.{\BBCQ}
\newblock
\APACjournalVolNumPages{Journal of Geophysical Research: Solid Earth}{127}{3}{e2021JB023598}.
\newblock
\begin{APACrefDOI} \doi{10.1029/2021JB023598} \end{APACrefDOI}
\PrintBackRefs{\CurrentBib}

\bibitem [\protect \citeauthoryear {%
Chen%
, Xia%
, Feng%
, Pang%
\BCBL {}\ \BBA {} Zhang%
}{%
Chen%
\ \protect \BOthers {.}}{%
{\protect \APACyear {2024}}%
}]{%
chen_2024_Surface}
\APACinsertmetastar {%
chen_2024_Surface}%
\begin{APACrefauthors}%
Chen, X.%
, Xia, J.%
, Feng, J.%
, Pang, J.%
\BCBL {}\ \BBA {} Zhang, H.%
\end{APACrefauthors}%
\unskip\
\newblock
\APACrefYearMonthDay{2024}{}{}.
\newblock
{\BBOQ}\APACrefatitle {Surface Wave Inversion Using a Multi-Information Fusion Neural Network} {Surface wave inversion using a multi-information fusion neural network}.{\BBCQ}
\newblock
\APACjournalVolNumPages{IEEE Transactions on Geoscience and Remote Sensing}{62}{}{1--13}.
\newblock
\begin{APACrefDOI} \doi{10.1109/TGRS.2024.3356663} \end{APACrefDOI}
\PrintBackRefs{\CurrentBib}

\bibitem [\protect \citeauthoryear {%
Chen%
, Xia%
, Pang%
, Zhou%
\BCBL {}\ \BBA {} Mi%
}{%
Chen%
\ \protect \BOthers {.}}{%
{\protect \APACyear {2022}}%
}]{%
chen_2022_Deep}
\APACinsertmetastar {%
chen_2022_Deep}%
\begin{APACrefauthors}%
Chen, X.%
, Xia, J.%
, Pang, J.%
, Zhou, C.%
\BCBL {}\ \BBA {} Mi, B.%
\end{APACrefauthors}%
\unskip\
\newblock
\APACrefYearMonthDay{2022}{}{}.
\newblock
{\BBOQ}\APACrefatitle {Deep Learning Inversion of Rayleigh-Wave Dispersion Curves with Geological Constraints for near-Surface Investigations} {Deep learning inversion of rayleigh-wave dispersion curves with geological constraints for near-surface investigations}.{\BBCQ}
\newblock
\APACjournalVolNumPages{Geophysical Journal International}{231}{1}{1--14}.
\newblock
\begin{APACrefDOI} \doi{10.1093/gji/ggac171} \end{APACrefDOI}
\PrintBackRefs{\CurrentBib}

\bibitem [\protect \citeauthoryear {%
Colombo%
, Turkoglu%
, Li%
, {Sandoval-Curiel}%
\BCBL {}\ \BBA {} Rovetta%
}{%
Colombo%
\ \protect \BOthers {.}}{%
{\protect \APACyear {2021}}%
}]{%
colombo_2021_Physicsdriven}
\APACinsertmetastar {%
colombo_2021_Physicsdriven}%
\begin{APACrefauthors}%
Colombo, D.%
, Turkoglu, E.%
, Li, W.%
, {Sandoval-Curiel}, E.%
\BCBL {}\ \BBA {} Rovetta, D.%
\end{APACrefauthors}%
\unskip\
\newblock
\APACrefYearMonthDay{2021}{}{}.
\newblock
{\BBOQ}\APACrefatitle {Physics-Driven Deep-Learning Inversion with Application to Transient Electromagnetics} {Physics-driven deep-learning inversion with application to transient electromagnetics}.{\BBCQ}
\newblock
\APACjournalVolNumPages{GEOPHYSICS}{86}{3}{E209-E224}.
\newblock
\begin{APACrefDOI} \doi{10.1190/geo2020-0760.1} \end{APACrefDOI}
\PrintBackRefs{\CurrentBib}

\bibitem [\protect \citeauthoryear {%
Dal~Moro%
\ \BBA {} Pipan%
}{%
Dal~Moro%
\ \BBA {} Pipan%
}{%
{\protect \APACyear {2007}}%
}]{%
dalmoro_2007_Joint}
\APACinsertmetastar {%
dalmoro_2007_Joint}%
\begin{APACrefauthors}%
Dal~Moro, G.%
\BCBT {}\ \BBA {} Pipan, M.%
\end{APACrefauthors}%
\unskip\
\newblock
\APACrefYearMonthDay{2007}{}{}.
\newblock
{\BBOQ}\APACrefatitle {Joint Inversion of Surface Wave Dispersion Curves and Reflection Travel Times via Multi-Objective Evolutionary Algorithms} {Joint inversion of surface wave dispersion curves and reflection travel times via multi-objective evolutionary algorithms}.{\BBCQ}
\newblock
\APACjournalVolNumPages{Journal of Applied Geophysics}{61}{1}{56--81}.
\newblock
\begin{APACrefDOI} \doi{10.1016/j.jappgeo.2006.04.001} \end{APACrefDOI}
\PrintBackRefs{\CurrentBib}

\bibitem [\protect \citeauthoryear {%
Deng%
\ \protect \BOthers {.}}{%
Deng%
\ \protect \BOthers {.}}{%
{\protect \APACyear {2022}}%
}]{%
deng_2022_Extended}
\APACinsertmetastar {%
deng_2022_Extended}%
\begin{APACrefauthors}%
Deng, B.%
, Li, J.%
, Liu, J.%
, Shen, C.%
, Suwen, J.%
\BCBL {}\ \BBA {} Chen, Q\BHBI F.%
\end{APACrefauthors}%
\unskip\
\newblock
\APACrefYearMonthDay{2022}{}{}.
\newblock
{\BBOQ}\APACrefatitle {The Extended Range Phase Shift Method for Broadband Surface Wave Dispersion Measurement from Ambient Noise and Its Application in Ore Deposit Characterization} {The extended range phase shift method for broadband surface wave dispersion measurement from ambient noise and its application in ore deposit characterization}.{\BBCQ}
\newblock
\APACjournalVolNumPages{Geophysics}{87}{3}{JM29-JM40}.
\newblock
\begin{APACrefDOI} \doi{10.1190/geo2021-0320.1} \end{APACrefDOI}
\PrintBackRefs{\CurrentBib}

\bibitem [\protect \citeauthoryear {%
Devilee%
, Curtis%
\BCBL {}\ \BBA {} Roy-Chowdhury%
}{%
Devilee%
\ \protect \BOthers {.}}{%
{\protect \APACyear {1999}}%
}]{%
devilee_1999_Efficient}
\APACinsertmetastar {%
devilee_1999_Efficient}%
\begin{APACrefauthors}%
Devilee, R\BPBI J\BPBI R.%
, Curtis, A.%
\BCBL {}\ \BBA {} Roy-Chowdhury, K.%
\end{APACrefauthors}%
\unskip\
\newblock
\APACrefYearMonthDay{1999}{}{}.
\newblock
{\BBOQ}\APACrefatitle {An Efficient, Probabilistic Neural Network Approach to Solving Inverse Problems: Inverting Surface Wave Velocities for Eurasian Crustal Thickness} {An efficient, probabilistic neural network approach to solving inverse problems: Inverting surface wave velocities for eurasian crustal thickness}.{\BBCQ}
\newblock
\APACjournalVolNumPages{Journal of Geophysical Research: Solid Earth}{104}{B12}{28841--28857}.
\newblock
\begin{APACrefDOI} \doi{10.1029/1999JB900273} \end{APACrefDOI}
\PrintBackRefs{\CurrentBib}

\bibitem [\protect \citeauthoryear {%
Earp%
, Curtis%
, Zhang%
\BCBL {}\ \BBA {} Hansteen%
}{%
Earp%
\ \protect \BOthers {.}}{%
{\protect \APACyear {2020}}%
}]{%
earp_2020_Probabilistic}
\APACinsertmetastar {%
earp_2020_Probabilistic}%
\begin{APACrefauthors}%
Earp, S.%
, Curtis, A.%
, Zhang, X.%
\BCBL {}\ \BBA {} Hansteen, F.%
\end{APACrefauthors}%
\unskip\
\newblock
\APACrefYearMonthDay{2020}{}{}.
\newblock
{\BBOQ}\APACrefatitle {Probabilistic Neural Network Tomography across Grane Field (North Sea) from Surface Wave Dispersion Data} {Probabilistic neural network tomography across grane field (north sea) from surface wave dispersion data}.{\BBCQ}
\newblock
\APACjournalVolNumPages{Geophysical Journal International}{223}{3}{1741--1757}.
\newblock
\begin{APACrefDOI} \doi{10.1093/gji/ggaa328} \end{APACrefDOI}
\PrintBackRefs{\CurrentBib}

\bibitem [\protect \citeauthoryear {%
Gan%
\ \protect \BOthers {.}}{%
Gan%
\ \protect \BOthers {.}}{%
{\protect \APACyear {2024}}%
}]{%
gan_2024_Deep}
\APACinsertmetastar {%
gan_2024_Deep}%
\begin{APACrefauthors}%
Gan, Y.%
, Yang, Z.%
, Pan, L.%
, Sun, Y\BHBI C.%
, Zhang, D.%
, Gao, Y.%
\BCBL {}\ \BBA {} Chen, X.%
\end{APACrefauthors}%
\unskip\
\newblock
\APACrefYearMonthDay{2024}{}{}.
\newblock
{\BBOQ}\APACrefatitle {Deep Learning-Based Dispersion Spectrum Inversion for Surface Wave Exploration} {Deep learning-based dispersion spectrum inversion for surface wave exploration}.{\BBCQ}
\newblock
\APACjournalVolNumPages{IEEE Transactions on Geoscience and Remote Sensing}{62}{}{1--11}.
\newblock
\begin{APACrefDOI} \doi{10.1109/TGRS.2024.3399033} \end{APACrefDOI}
\PrintBackRefs{\CurrentBib}

\bibitem [\protect \citeauthoryear {%
Gouveia%
\ \BBA {} Scales%
}{%
Gouveia%
\ \BBA {} Scales%
}{%
{\protect \APACyear {1998}}%
}]{%
gouveia_1998_Bayesian}
\APACinsertmetastar {%
gouveia_1998_Bayesian}%
\begin{APACrefauthors}%
Gouveia, W\BPBI P.%
\BCBT {}\ \BBA {} Scales, J\BPBI A.%
\end{APACrefauthors}%
\unskip\
\newblock
\APACrefYearMonthDay{1998}{}{}.
\newblock
{\BBOQ}\APACrefatitle {Bayesian Seismic Waveform Inversion: Parameter Estimation and Uncertainty Analysis} {Bayesian seismic waveform inversion: Parameter estimation and uncertainty analysis}.{\BBCQ}
\newblock
\APACjournalVolNumPages{Journal of Geophysical Research: Solid Earth}{103}{B2}{2759--2779}.
\newblock
\begin{APACrefDOI} \doi{10.1029/97JB02933} \end{APACrefDOI}
\PrintBackRefs{\CurrentBib}

\bibitem [\protect \citeauthoryear {%
Griffiths%
, Cox%
, Rathje%
\BCBL {}\ \BBA {} Teague%
}{%
Griffiths%
\ \protect \BOthers {.}}{%
{\protect \APACyear {2016}}%
}]{%
griffiths_2016_Surfacewave}
\APACinsertmetastar {%
griffiths_2016_Surfacewave}%
\begin{APACrefauthors}%
Griffiths, S\BPBI C.%
, Cox, B\BPBI R.%
, Rathje, E\BPBI M.%
\BCBL {}\ \BBA {} Teague, D\BPBI P.%
\end{APACrefauthors}%
\unskip\
\newblock
\APACrefYearMonthDay{2016}{}{}.
\newblock
{\BBOQ}\APACrefatitle {Surface-Wave Dispersion Approach for Evaluating Statistical Models That Account for Shear-Wave Velocity Uncertainty} {Surface-wave dispersion approach for evaluating statistical models that account for shear-wave velocity uncertainty}.{\BBCQ}
\newblock
\APACjournalVolNumPages{Journal of Geotechnical and Geoenvironmental Engineering}{142}{11}{4016061}.
\newblock
\begin{APACrefDOI} \doi{10.1061/(ASCE)GT.1943-5606.0001552} \end{APACrefDOI}
\PrintBackRefs{\CurrentBib}

\bibitem [\protect \citeauthoryear {%
Haskell%
}{%
Haskell%
}{%
{\protect \APACyear {1953}}%
}]{%
haskell_1953_Dispersion}
\APACinsertmetastar {%
haskell_1953_Dispersion}%
\begin{APACrefauthors}%
Haskell, N\BPBI A.%
\end{APACrefauthors}%
\unskip\
\newblock
\APACrefYearMonthDay{1953}{}{}.
\newblock
{\BBOQ}\APACrefatitle {The Dispersion of Surface Waves on Multilayered Media} {The dispersion of surface waves on multilayered media}.{\BBCQ}
\newblock
\BIn{} A.~Ben-Menahem\ (\BED), \APACrefbtitle {Vincit Veritas: A Portrait of the Life and Work of Norman Abraham Haskell, 1905--1970} {Vincit veritas: A portrait of the life and work of norman abraham haskell, 1905--1970}\ (\BVOL~43, \BPGS\ 86--103).
\newblock
\APACaddressPublisher{Washington, D. C.}{American Geophysical Union}.
\newblock
\begin{APACrefDOI} \doi{10.1785/BSSA0430010017} \end{APACrefDOI}
\PrintBackRefs{\CurrentBib}

\bibitem [\protect \citeauthoryear {%
Herrmann%
}{%
Herrmann%
}{%
{\protect \APACyear {2013}}%
}]{%
herrmann_2013_Computer}
\APACinsertmetastar {%
herrmann_2013_Computer}%
\begin{APACrefauthors}%
Herrmann, R\BPBI B.%
\end{APACrefauthors}%
\unskip\
\newblock
\APACrefYearMonthDay{2013}{}{}.
\newblock
{\BBOQ}\APACrefatitle {Computer Programs in Seismology: An Evolving Tool for Instruction and Research} {Computer programs in seismology: An evolving tool for instruction and research}.{\BBCQ}
\newblock
\APACjournalVolNumPages{Seismological Research Letters}{84}{6}{1081--1088}.
\newblock
\begin{APACrefDOI} \doi{10.1785/0220110096} \end{APACrefDOI}
\PrintBackRefs{\CurrentBib}

\bibitem [\protect \citeauthoryear {%
Herrmann%
, Ammon%
, Benz%
, {Aziz-Zanjani}%
\BCBL {}\ \BBA {} Boschelli%
}{%
Herrmann%
\ \protect \BOthers {.}}{%
{\protect \APACyear {2021}}%
}]{%
herrmann_2021_ShortPeriod}
\APACinsertmetastar {%
herrmann_2021_ShortPeriod}%
\begin{APACrefauthors}%
Herrmann, R\BPBI B.%
, Ammon, C\BPBI J.%
, Benz, H\BPBI M.%
, {Aziz-Zanjani}, A.%
\BCBL {}\ \BBA {} Boschelli, J.%
\end{APACrefauthors}%
\unskip\
\newblock
\APACrefYearMonthDay{2021}{}{}.
\newblock
{\BBOQ}\APACrefatitle {Short-Period Surface-Wave Tomography in the Continental United States---A Resource for Research} {Short-period surface-wave tomography in the continental united states---a resource for research}.{\BBCQ}
\newblock
\APACjournalVolNumPages{Seismological Research Letters}{92}{6}{3642--3656}.
\newblock
\begin{APACrefDOI} \doi{10.1785/0220200462} \end{APACrefDOI}
\PrintBackRefs{\CurrentBib}

\bibitem [\protect \citeauthoryear {%
Hu%
, Qiu%
, Zhang%
\BCBL {}\ \BBA {} {Ben-Zion}%
}{%
Hu%
\ \protect \BOthers {.}}{%
{\protect \APACyear {2020}}%
}]{%
hu_2020_Using}
\APACinsertmetastar {%
hu_2020_Using}%
\begin{APACrefauthors}%
Hu, J.%
, Qiu, H.%
, Zhang, H.%
\BCBL {}\ \BBA {} {Ben-Zion}, Y.%
\end{APACrefauthors}%
\unskip\
\newblock
\APACrefYearMonthDay{2020}{}{}.
\newblock
{\BBOQ}\APACrefatitle {Using Deep Learning to Derive Shear-Wave Velocity Models from Surface-Wave Dispersion Data} {Using deep learning to derive shear-wave velocity models from surface-wave dispersion data}.{\BBCQ}
\newblock
\APACjournalVolNumPages{Seismological Research Letters}{91}{3}{1738--1751}.
\newblock
\begin{APACrefDOI} \doi{10.1785/0220190222} \end{APACrefDOI}
\PrintBackRefs{\CurrentBib}

\bibitem [\protect \citeauthoryear {%
Kaviani%
\ \protect \BOthers {.}}{%
Kaviani%
\ \protect \BOthers {.}}{%
{\protect \APACyear {2020}}%
}]{%
kaviani_2020_Crustal}
\APACinsertmetastar {%
kaviani_2020_Crustal}%
\begin{APACrefauthors}%
Kaviani, A.%
, Paul, A.%
, Moradi, A.%
, Mai, P\BPBI M.%
, Pilia, S.%
, Boschi, L.%
\BDBL {}Sandvol, E.%
\end{APACrefauthors}%
\unskip\
\newblock
\APACrefYearMonthDay{2020}{}{}.
\newblock
{\BBOQ}\APACrefatitle {Crustal and Uppermost Mantle Shear Wave Velocity Structure beneath the Middle East from Surface Wave Tomography} {Crustal and uppermost mantle shear wave velocity structure beneath the middle east from surface wave tomography}.{\BBCQ}
\newblock
\APACjournalVolNumPages{Geophysical Journal International}{221}{2}{1349--1365}.
\newblock
\begin{APACrefDOI} \doi{10.1093/gji/ggaa075} \end{APACrefDOI}
\PrintBackRefs{\CurrentBib}

\bibitem [\protect \citeauthoryear {%
Kennett%
, Engdahl%
\BCBL {}\ \BBA {} Buland%
}{%
Kennett%
\ \protect \BOthers {.}}{%
{\protect \APACyear {1995}}%
}]{%
kennett_1995_Constraints}
\APACinsertmetastar {%
kennett_1995_Constraints}%
\begin{APACrefauthors}%
Kennett, B\BPBI L\BPBI N.%
, Engdahl, E\BPBI R.%
\BCBL {}\ \BBA {} Buland, R.%
\end{APACrefauthors}%
\unskip\
\newblock
\APACrefYearMonthDay{1995}{}{}.
\newblock
{\BBOQ}\APACrefatitle {Constraints on Seismic Velocities in the Earth from Traveltimes} {Constraints on seismic velocities in the earth from traveltimes}.{\BBCQ}
\newblock
\APACjournalVolNumPages{Geophysical Journal International}{122}{1}{108--124}.
\newblock
\begin{APACrefDOI} \doi{10.1111/j.1365-246X.1995.tb03540.x} \end{APACrefDOI}
\PrintBackRefs{\CurrentBib}

\bibitem [\protect \citeauthoryear {%
Kingma%
\ \BBA {} Ba%
}{%
Kingma%
\ \BBA {} Ba%
}{%
{\protect \APACyear {2017}}%
}]{%
kingma_2017_Adam}
\APACinsertmetastar {%
kingma_2017_Adam}%
\begin{APACrefauthors}%
Kingma, D\BPBI P.%
\BCBT {}\ \BBA {} Ba, J.%
\end{APACrefauthors}%
\unskip\
\newblock
\APACrefYearMonthDay{2017}{}{}.
\newblock
\APACrefbtitle {Adam: A Method for Stochastic Optimization} {Adam: A method for stochastic optimization}\ (\BNUM\ arXiv:1412.6980).
\newblock
\APACaddressPublisher{}{arXiv}.
\newblock
\begin{APACrefDOI} \doi{10.48550/arXiv.1412.6980} \end{APACrefDOI}
\PrintBackRefs{\CurrentBib}

\bibitem [\protect \citeauthoryear {%
Lai%
, Foti%
\BCBL {}\ \BBA {} Rix%
}{%
Lai%
\ \protect \BOthers {.}}{%
{\protect \APACyear {2005}}%
}]{%
lai_2005_Propagation}
\APACinsertmetastar {%
lai_2005_Propagation}%
\begin{APACrefauthors}%
Lai, C\BPBI G.%
, Foti, S.%
\BCBL {}\ \BBA {} Rix, G\BPBI J.%
\end{APACrefauthors}%
\unskip\
\newblock
\APACrefYearMonthDay{2005}{}{}.
\newblock
{\BBOQ}\APACrefatitle {Propagation of Data Uncertainty in Surface Wave Inversion} {Propagation of data uncertainty in surface wave inversion}.{\BBCQ}
\newblock
\APACjournalVolNumPages{Journal of Environmental and Engineering Geophysics}{10}{2}{219--228}.
\newblock
\begin{APACrefDOI} \doi{10.2113/JEEG10.2.219} \end{APACrefDOI}
\PrintBackRefs{\CurrentBib}

\bibitem [\protect \citeauthoryear {%
A.~Li%
\ \BBA {} Li%
}{%
A.~Li%
\ \BBA {} Li%
}{%
{\protect \APACyear {2015}}%
}]{%
li_2015_Love}
\APACinsertmetastar {%
li_2015_Love}%
\begin{APACrefauthors}%
Li, A.%
\BCBT {}\ \BBA {} Li, L.%
\end{APACrefauthors}%
\unskip\
\newblock
\APACrefYearMonthDay{2015}{}{}.
\newblock
{\BBOQ}\APACrefatitle {Love Wave Tomography in Southern Africa from a Two-Plane-Wave Inversion Method} {Love wave tomography in southern africa from a two-plane-wave inversion method}.{\BBCQ}
\newblock
\APACjournalVolNumPages{Geophysical Journal International}{202}{2}{1005--1020}.
\newblock
\begin{APACrefDOI} \doi{10.1093/gji/ggv203} \end{APACrefDOI}
\PrintBackRefs{\CurrentBib}

\bibitem [\protect \citeauthoryear {%
C.~Li%
, Liu%
\BCBL {}\ \protect \BOthers {.}}{%
C.~Li%
, Liu%
\BCBL {}\ \protect \BOthers {.}}{%
{\protect \APACyear {2025}}%
}]{%
li_2025_Simultaneous}
\APACinsertmetastar {%
li_2025_Simultaneous}%
\begin{APACrefauthors}%
Li, C.%
, Liu, G.%
, Wang, Z.%
, Li, Z.%
, Fomel, S.%
\BCBL {}\ \BBA {} Chen, Y.%
\end{APACrefauthors}%
\unskip\
\newblock
\APACrefYearMonthDay{2025}{}{}.
\newblock
{\BBOQ}\APACrefatitle {Simultaneous Off-the-Grid Deblending and Data Reconstruction via Unsupervised Deep Learning} {Simultaneous off-the-grid deblending and data reconstruction via unsupervised deep learning}.{\BBCQ}
\newblock
\APACjournalVolNumPages{IEEE Transactions on Geoscience and Remote Sensing}{63}{}{1--11}.
\newblock
\begin{APACrefDOI} \doi{10.1109/TGRS.2025.3548644} \end{APACrefDOI}
\PrintBackRefs{\CurrentBib}

\bibitem [\protect \citeauthoryear {%
C.~Li%
, Saad%
\BCBL {}\ \BBA {} Chen%
}{%
C.~Li%
, Saad%
\BCBL {}\ \BBA {} Chen%
}{%
{\protect \APACyear {2025}}%
}]{%
li_2025_Unsupervised}
\APACinsertmetastar {%
li_2025_Unsupervised}%
\begin{APACrefauthors}%
Li, C.%
, Saad, O\BPBI M.%
\BCBL {}\ \BBA {} Chen, Y.%
\end{APACrefauthors}%
\unskip\
\newblock
\APACrefYearMonthDay{2025}{}{}.
\newblock
{\BBOQ}\APACrefatitle {Unsupervised Deep Learning for Off-the-Grid Seismic Reconstruction and Denoising} {Unsupervised deep learning for off-the-grid seismic reconstruction and denoising}.{\BBCQ}
\newblock
\APACjournalVolNumPages{Geophysics}{90}{3}{V241-V254}.
\newblock
\begin{APACrefDOI} \doi{10.1190/geo2024-0189.1} \end{APACrefDOI}
\PrintBackRefs{\CurrentBib}

\bibitem [\protect \citeauthoryear {%
S.~Li%
\ \protect \BOthers {.}}{%
S.~Li%
\ \protect \BOthers {.}}{%
{\protect \APACyear {2020}}%
}]{%
li_2020_DeepLearning}
\APACinsertmetastar {%
li_2020_DeepLearning}%
\begin{APACrefauthors}%
Li, S.%
, Liu, B.%
, Ren, Y.%
, Chen, Y.%
, Yang, S.%
, Wang, Y.%
\BCBL {}\ \BBA {} Jiang, P.%
\end{APACrefauthors}%
\unskip\
\newblock
\APACrefYearMonthDay{2020}{}{}.
\newblock
{\BBOQ}\APACrefatitle {Deep-Learning Inversion of Seismic Data} {Deep-learning inversion of seismic data}.{\BBCQ}
\newblock
\APACjournalVolNumPages{IEEE Transactions on Geoscience and Remote Sensing}{58}{3}{2135--2149}.
\newblock
\begin{APACrefDOI} \doi{10.1109/TGRS.2019.2953473} \end{APACrefDOI}
\PrintBackRefs{\CurrentBib}

\bibitem [\protect \citeauthoryear {%
Y.~Li%
, Jia%
\BCBL {}\ \BBA {} Lu%
}{%
Y.~Li%
\ \protect \BOthers {.}}{%
{\protect \APACyear {2022}}%
}]{%
li_2022_SelfSupervised}
\APACinsertmetastar {%
li_2022_SelfSupervised}%
\begin{APACrefauthors}%
Li, Y.%
, Jia, Z.%
\BCBL {}\ \BBA {} Lu, W.%
\end{APACrefauthors}%
\unskip\
\newblock
\APACrefYearMonthDay{2022}{}{}.
\newblock
{\BBOQ}\APACrefatitle {Self-Supervised Deep Learning for 3D Gravity Inversion} {Self-supervised deep learning for 3d gravity inversion}.{\BBCQ}
\newblock
\APACjournalVolNumPages{IEEE Transactions on Geoscience and Remote Sensing}{60}{}{1--11}.
\newblock
\begin{APACrefDOI} \doi{10.1109/TGRS.2022.3225449} \end{APACrefDOI}
\PrintBackRefs{\CurrentBib}

\bibitem [\protect \citeauthoryear {%
F.~Liu%
}{%
F.~Liu%
}{%
{\protect \APACyear {2025}}%
}]{%
liu_2025_Surface}
\APACinsertmetastar {%
liu_2025_Surface}%
\begin{APACrefauthors}%
Liu, F.%
\end{APACrefauthors}%
\unskip\
\newblock
\APACrefYearMonthDay{2025}{}{}.
\newblock
\APACrefbtitle {Surface Wave Dispersion Benchmark Datasets: Synthetic and Real-World Cases: DispFormer Training Datasets.} {Surface wave dispersion benchmark datasets: Synthetic and real-world cases: Dispformer training datasets.}
\newblock
\APACaddressPublisher{}{Zenodo}.
\newblock
\begin{APACrefDOI} \doi{10.5281/ZENODO.14619577} \end{APACrefDOI}
\PrintBackRefs{\CurrentBib}

\bibitem [\protect \citeauthoryear {%
F.~Liu%
, Li%
, Fu%
\BCBL {}\ \BBA {} Lu%
}{%
F.~Liu%
\ \protect \BOthers {.}}{%
{\protect \APACyear {2024}}%
}]{%
liu_2024_Multimodal}
\APACinsertmetastar {%
liu_2024_Multimodal}%
\begin{APACrefauthors}%
Liu, F.%
, Li, J.%
, Fu, L.%
\BCBL {}\ \BBA {} Lu, L.%
\end{APACrefauthors}%
\unskip\
\newblock
\APACrefYearMonthDay{2024}{}{}.
\newblock
{\BBOQ}\APACrefatitle {Multimodal Surface Wave Inversion with Automatic Differentiation} {Multimodal surface wave inversion with automatic differentiation}.{\BBCQ}
\newblock
\APACjournalVolNumPages{Geophysical Journal International}{238}{1}{290--312}.
\newblock
\begin{APACrefDOI} \doi{10.1093/gji/ggae155} \end{APACrefDOI}
\PrintBackRefs{\CurrentBib}

\bibitem [\protect \citeauthoryear {%
W.~Liu%
, Wang%
, Xi%
, Zhang%
\BCBL {}\ \BBA {} Huang%
}{%
W.~Liu%
\ \protect \BOthers {.}}{%
{\protect \APACyear {2022}}%
}]{%
liu_2022_PhysicsDriven}
\APACinsertmetastar {%
liu_2022_PhysicsDriven}%
\begin{APACrefauthors}%
Liu, W.%
, Wang, H.%
, Xi, Z.%
, Zhang, R.%
\BCBL {}\ \BBA {} Huang, X.%
\end{APACrefauthors}%
\unskip\
\newblock
\APACrefYearMonthDay{2022}{}{}.
\newblock
{\BBOQ}\APACrefatitle {Physics-Driven Deep Learning Inversion with Application to Magnetotelluric} {Physics-driven deep learning inversion with application to magnetotelluric}.{\BBCQ}
\newblock
\APACjournalVolNumPages{Remote Sensing}{14}{13}{3218}.
\newblock
\begin{APACrefDOI} \doi{10.3390/rs14133218} \end{APACrefDOI}
\PrintBackRefs{\CurrentBib}

\bibitem [\protect \citeauthoryear {%
Y.~Liu%
\ \protect \BOthers {.}}{%
Y.~Liu%
\ \protect \BOthers {.}}{%
{\protect \APACyear {2023}}%
}]{%
liu_2023_Highresolution}
\APACinsertmetastar {%
liu_2023_Highresolution}%
\begin{APACrefauthors}%
Liu, Y.%
, Yu, Z.%
, Zhang, Z.%
, Yao, H.%
, Wang, W.%
, Zhang, H.%
\BDBL {}Fang, L.%
\end{APACrefauthors}%
\unskip\
\newblock
\APACrefYearMonthDay{2023}{}{}.
\newblock
{\BBOQ}\APACrefatitle {The High-Resolution Community Velocity Model V2.0 of Southwest China, Constructed by Joint Body and Surface Wave Tomography of Data Recorded at Temporary Dense Arrays} {The high-resolution community velocity model v2.0 of southwest china, constructed by joint body and surface wave tomography of data recorded at temporary dense arrays}.{\BBCQ}
\newblock
\APACjournalVolNumPages{Science China Earth Sciences}{66}{10}{2368--2385}.
\newblock
\begin{APACrefDOI} \doi{10.1007/s11430-022-1161-7} \end{APACrefDOI}
\PrintBackRefs{\CurrentBib}

\bibitem [\protect \citeauthoryear {%
Luo%
\ \protect \BOthers {.}}{%
Luo%
\ \protect \BOthers {.}}{%
{\protect \APACyear {2022}}%
}]{%
luo_2022_Constructing}
\APACinsertmetastar {%
luo_2022_Constructing}%
\begin{APACrefauthors}%
Luo, Y.%
, Huang, Y.%
, Yang, Y.%
, Zhao, K.%
, Yang, X.%
\BCBL {}\ \BBA {} Xu, H.%
\end{APACrefauthors}%
\unskip\
\newblock
\APACrefYearMonthDay{2022}{}{}.
\newblock
{\BBOQ}\APACrefatitle {Constructing Shear Velocity Models from Surface Wave Dispersion Curves Using Deep Learning} {Constructing shear velocity models from surface wave dispersion curves using deep learning}.{\BBCQ}
\newblock
\APACjournalVolNumPages{Journal of Applied Geophysics}{196}{}{104524}.
\newblock
\begin{APACrefDOI} \doi{10.1016/j.jappgeo.2021.104524} \end{APACrefDOI}
\PrintBackRefs{\CurrentBib}

\bibitem [\protect \citeauthoryear {%
Luu%
}{%
Luu%
}{%
{\protect \APACyear {2023}}%
}]{%
luu_2023_Evodcinv}
\APACinsertmetastar {%
luu_2023_Evodcinv}%
\begin{APACrefauthors}%
Luu, K.%
\end{APACrefauthors}%
\unskip\
\newblock
\APACrefYearMonthDay{2023}{}{}.
\newblock
\APACrefbtitle {Evodcinv: Inversion of Dispersion Curves Using Evolutionary Algorithms.} {Evodcinv: Inversion of dispersion curves using evolutionary algorithms.}
\newblock
\APAChowpublished {Zenodo}.
\newblock
\begin{APACrefDOI} \doi{10.5281/ZENODO.10112876} \end{APACrefDOI}
\PrintBackRefs{\CurrentBib}

\bibitem [\protect \citeauthoryear {%
Maraschini%
\ \BBA {} Foti%
}{%
Maraschini%
\ \BBA {} Foti%
}{%
{\protect \APACyear {2010}}%
}]{%
maraschini_2010_Monte}
\APACinsertmetastar {%
maraschini_2010_Monte}%
\begin{APACrefauthors}%
Maraschini, M.%
\BCBT {}\ \BBA {} Foti, S.%
\end{APACrefauthors}%
\unskip\
\newblock
\APACrefYearMonthDay{2010}{}{}.
\newblock
{\BBOQ}\APACrefatitle {A Monte Carlo Multimodal Inversion of Surface Waves: Monte Carlo Multimodal Surface Wave Inversion} {A monte carlo multimodal inversion of surface waves: Monte carlo multimodal surface wave inversion}.{\BBCQ}
\newblock
\APACjournalVolNumPages{Geophysical Journal International}{182}{3}{1557--1566}.
\newblock
\begin{APACrefDOI} \doi{10.1111/j.1365-246X.2010.04703.x} \end{APACrefDOI}
\PrintBackRefs{\CurrentBib}

\bibitem [\protect \citeauthoryear {%
R.~Meier%
\ \BBA {} Rix%
}{%
R.~Meier%
\ \BBA {} Rix%
}{%
{\protect \APACyear {1993}}%
}]{%
meier_1993_Initial}
\APACinsertmetastar {%
meier_1993_Initial}%
\begin{APACrefauthors}%
Meier, R.%
\BCBT {}\ \BBA {} Rix, G.%
\end{APACrefauthors}%
\unskip\
\newblock
\APACrefYearMonthDay{1993}{}{}.
\newblock
{\BBOQ}\APACrefatitle {An Initial Study of Surface Wave Inversion Using Artificial Neural Networks} {An initial study of surface wave inversion using artificial neural networks}.{\BBCQ}
\newblock
\APACjournalVolNumPages{Geotechnical Testing Journal}{16}{4}{425--431}.
\newblock
\begin{APACrefDOI} \doi{10.1520/GTJ10282J} \end{APACrefDOI}
\PrintBackRefs{\CurrentBib}

\bibitem [\protect \citeauthoryear {%
U.~Meier%
, Curtis%
\BCBL {}\ \BBA {} Trampert%
}{%
U.~Meier%
\ \protect \BOthers {.}}{%
{\protect \APACyear {2007}}%
}]{%
meier_2007_Global}
\APACinsertmetastar {%
meier_2007_Global}%
\begin{APACrefauthors}%
Meier, U.%
, Curtis, A.%
\BCBL {}\ \BBA {} Trampert, J.%
\end{APACrefauthors}%
\unskip\
\newblock
\APACrefYearMonthDay{2007}{}{}.
\newblock
{\BBOQ}\APACrefatitle {Global Crustal Thickness from Neural Network Inversion of Surface Wave Data} {Global crustal thickness from neural network inversion of surface wave data}.{\BBCQ}
\newblock
\APACjournalVolNumPages{Geophysical Journal International}{169}{2}{706--722}.
\newblock
\begin{APACrefDOI} \doi{10.1111/j.1365-246X.2007.03373.x} \end{APACrefDOI}
\PrintBackRefs{\CurrentBib}

\bibitem [\protect \citeauthoryear {%
Montagner%
\ \BBA {} Tanimoto%
}{%
Montagner%
\ \BBA {} Tanimoto%
}{%
{\protect \APACyear {1990}}%
}]{%
montagner_1990_Global}
\APACinsertmetastar {%
montagner_1990_Global}%
\begin{APACrefauthors}%
Montagner, J\BHBI P.%
\BCBT {}\ \BBA {} Tanimoto, T.%
\end{APACrefauthors}%
\unskip\
\newblock
\APACrefYearMonthDay{1990}{}{}.
\newblock
{\BBOQ}\APACrefatitle {Global Anisotropy in the Upper Mantle Inferred from the Regionalization of Phase Velocities} {Global anisotropy in the upper mantle inferred from the regionalization of phase velocities}.{\BBCQ}
\newblock
\APACjournalVolNumPages{Journal of Geophysical Research: Solid Earth}{95}{B4}{4797--4819}.
\newblock
\begin{APACrefDOI} \doi{10.1029/JB095iB04p04797} \end{APACrefDOI}
\PrintBackRefs{\CurrentBib}

\bibitem [\protect \citeauthoryear {%
Mordret%
\ \protect \BOthers {.}}{%
Mordret%
\ \protect \BOthers {.}}{%
{\protect \APACyear {2013}}%
}]{%
mordret_2013_Nearsurface}
\APACinsertmetastar {%
mordret_2013_Nearsurface}%
\begin{APACrefauthors}%
Mordret, A.%
, Land{\`e}s, M.%
, Shapiro, N\BPBI M.%
, Singh, S\BPBI C.%
, Roux, P.%
\BCBL {}\ \BBA {} Barkved, O\BPBI I.%
\end{APACrefauthors}%
\unskip\
\newblock
\APACrefYearMonthDay{2013}{}{}.
\newblock
{\BBOQ}\APACrefatitle {Near-Surface Study at the Valhall Oil Field from Ambient Noise Surface Wave Tomography} {Near-surface study at the valhall oil field from ambient noise surface wave tomography}.{\BBCQ}
\newblock
\APACjournalVolNumPages{Geophysical Journal International}{193}{3}{1627--1643}.
\newblock
\begin{APACrefDOI} \doi{10.1093/gji/ggt061} \end{APACrefDOI}
\PrintBackRefs{\CurrentBib}

\bibitem [\protect \citeauthoryear {%
Nishida%
, Montagner%
\BCBL {}\ \BBA {} Kawakatsu%
}{%
Nishida%
\ \protect \BOthers {.}}{%
{\protect \APACyear {2009}}%
}]{%
nishida_2009_Global}
\APACinsertmetastar {%
nishida_2009_Global}%
\begin{APACrefauthors}%
Nishida, K.%
, Montagner, J\BHBI P.%
\BCBL {}\ \BBA {} Kawakatsu, H.%
\end{APACrefauthors}%
\unskip\
\newblock
\APACrefYearMonthDay{2009}{}{}.
\newblock
{\BBOQ}\APACrefatitle {Global Surface Wave Tomography Using Seismic Hum} {Global surface wave tomography using seismic hum}.{\BBCQ}
\newblock
\APACjournalVolNumPages{Science}{326}{5949}{112--112}.
\newblock
\begin{APACrefDOI} \doi{10.1126/science.1176389} \end{APACrefDOI}
\PrintBackRefs{\CurrentBib}

\bibitem [\protect \citeauthoryear {%
Pasyanos%
, Masters%
, Laske%
\BCBL {}\ \BBA {} Ma%
}{%
Pasyanos%
\ \protect \BOthers {.}}{%
{\protect \APACyear {2014}}%
}]{%
pasyanos_2014_LITHO10}
\APACinsertmetastar {%
pasyanos_2014_LITHO10}%
\begin{APACrefauthors}%
Pasyanos, M\BPBI E.%
, Masters, T\BPBI G.%
, Laske, G.%
\BCBL {}\ \BBA {} Ma, Z.%
\end{APACrefauthors}%
\unskip\
\newblock
\APACrefYearMonthDay{2014}{}{}.
\newblock
{\BBOQ}\APACrefatitle {LITHO1.0: An Updated Crust and Lithospheric Model of the Earth} {Litho1.0: An updated crust and lithospheric model of the earth}.{\BBCQ}
\newblock
\APACjournalVolNumPages{Journal of Geophysical Research: Solid Earth}{119}{3}{2153--2173}.
\newblock
\begin{APACrefDOI} \doi{10.1002/2013JB010626} \end{APACrefDOI}
\PrintBackRefs{\CurrentBib}

\bibitem [\protect \citeauthoryear {%
Qin%
, Lu%
, Ding%
, Feng%
\BCBL {}\ \BBA {} Zhang%
}{%
Qin%
\ \protect \BOthers {.}}{%
{\protect \APACyear {2022}}%
}]{%
qin_2022_HighResolution}
\APACinsertmetastar {%
qin_2022_HighResolution}%
\begin{APACrefauthors}%
Qin, T.%
, Lu, L.%
, Ding, Z.%
, Feng, X.%
\BCBL {}\ \BBA {} Zhang, Y.%
\end{APACrefauthors}%
\unskip\
\newblock
\APACrefYearMonthDay{2022}{}{}.
\newblock
{\BBOQ}\APACrefatitle {High-Resolution 3D Shallow {\emph{S}} Wave Velocity Structure of Tongzhou, Subcenter of Beijing, Inferred From Multimode Rayleigh Waves by Beamforming Seismic Noise at a Dense Array} {High-resolution 3d shallow {\emph{s}} wave velocity structure of tongzhou, subcenter of beijing, inferred from multimode rayleigh waves by beamforming seismic noise at a dense array}.{\BBCQ}
\newblock
\APACjournalVolNumPages{Journal of Geophysical Research: Solid Earth}{127}{5}{e2021JB023689}.
\newblock
\begin{APACrefDOI} \doi{10.1029/2021JB023689} \end{APACrefDOI}
\PrintBackRefs{\CurrentBib}

\bibitem [\protect \citeauthoryear {%
Sager%
, Ermert%
, Boehm%
\BCBL {}\ \BBA {} Fichtner%
}{%
Sager%
\ \protect \BOthers {.}}{%
{\protect \APACyear {2018}}%
}]{%
sager_2018_Full}
\APACinsertmetastar {%
sager_2018_Full}%
\begin{APACrefauthors}%
Sager, K.%
, Ermert, L.%
, Boehm, C.%
\BCBL {}\ \BBA {} Fichtner, A.%
\end{APACrefauthors}%
\unskip\
\newblock
\APACrefYearMonthDay{2018}{}{}.
\newblock
{\BBOQ}\APACrefatitle {Towards Full Waveform Ambient Noise Inversion} {Towards full waveform ambient noise inversion}.{\BBCQ}
\newblock
\APACjournalVolNumPages{Geophysical Journal International}{212}{1}{566--590}.
\newblock
\begin{APACrefDOI} \doi{10.1093/gji/ggx429} \end{APACrefDOI}
\PrintBackRefs{\CurrentBib}

\bibitem [\protect \citeauthoryear {%
Sambridge%
\ \BBA {} Mosegaard%
}{%
Sambridge%
\ \BBA {} Mosegaard%
}{%
{\protect \APACyear {2002}}%
}]{%
sambridge_2002_MONTE}
\APACinsertmetastar {%
sambridge_2002_MONTE}%
\begin{APACrefauthors}%
Sambridge, M.%
\BCBT {}\ \BBA {} Mosegaard, K.%
\end{APACrefauthors}%
\unskip\
\newblock
\APACrefYearMonthDay{2002}{}{}.
\newblock
{\BBOQ}\APACrefatitle {MONTE CARLO METHODS IN GEOPHYSICAL INVERSE PROBLEMS} {Monte carlo methods in geophysical inverse problems}.{\BBCQ}
\newblock
\APACjournalVolNumPages{Reviews of Geophysics}{40}{3}{}.
\newblock
\begin{APACrefDOI} \doi{10.1029/2000RG000089} \end{APACrefDOI}
\PrintBackRefs{\CurrentBib}

\bibitem [\protect \citeauthoryear {%
Sen%
\ \BBA {} Stoffa%
}{%
Sen%
\ \BBA {} Stoffa%
}{%
{\protect \APACyear {2013}}%
}]{%
sen_2013_Global}
\APACinsertmetastar {%
sen_2013_Global}%
\begin{APACrefauthors}%
Sen, M\BPBI K.%
\BCBT {}\ \BBA {} Stoffa, P\BPBI L.%
\end{APACrefauthors}%
\unskip\
\newblock
\APACrefYear{2013}.
\newblock
\APACrefbtitle {Global Optimization Methods in Geophysical Inversion} {Global optimization methods in geophysical inversion}\ (\PrintOrdinal{2}\ \BEd).
\newblock
\APACaddressPublisher{}{Cambridge University Press}.
\newblock
\begin{APACrefDOI} \doi{10.1017/CBO9780511997570} \end{APACrefDOI}
\PrintBackRefs{\CurrentBib}

\bibitem [\protect \citeauthoryear {%
Shapiro%
, Campillo%
, Stehly%
\BCBL {}\ \BBA {} Ritzwoller%
}{%
Shapiro%
\ \protect \BOthers {.}}{%
{\protect \APACyear {2005}}%
}]{%
shapiro_2005_Highresolution}
\APACinsertmetastar {%
shapiro_2005_Highresolution}%
\begin{APACrefauthors}%
Shapiro, N\BPBI M.%
, Campillo, M.%
, Stehly, L.%
\BCBL {}\ \BBA {} Ritzwoller, M\BPBI H.%
\end{APACrefauthors}%
\unskip\
\newblock
\APACrefYearMonthDay{2005}{}{}.
\newblock
{\BBOQ}\APACrefatitle {High-Resolution Surface-Wave Tomography from Ambient Seismic Noise} {High-resolution surface-wave tomography from ambient seismic noise}.{\BBCQ}
\newblock
\APACjournalVolNumPages{Science}{307}{5715}{1615--1618}.
\newblock
\begin{APACrefDOI} \doi{10.1126/science.1108339} \end{APACrefDOI}
\PrintBackRefs{\CurrentBib}

\bibitem [\protect \citeauthoryear {%
Shapiro%
\ \BBA {} Ritzwoller%
}{%
Shapiro%
\ \BBA {} Ritzwoller%
}{%
{\protect \APACyear {2002}}%
}]{%
shapiro_2002_Montecarlo}
\APACinsertmetastar {%
shapiro_2002_Montecarlo}%
\begin{APACrefauthors}%
Shapiro, N\BPBI M.%
\BCBT {}\ \BBA {} Ritzwoller, M\BPBI H.%
\end{APACrefauthors}%
\unskip\
\newblock
\APACrefYearMonthDay{2002}{}{}.
\newblock
{\BBOQ}\APACrefatitle {Monte-Carlo Inversion for a Global Shear-Velocity Model of the Crust and Upper Mantle} {Monte-carlo inversion for a global shear-velocity model of the crust and upper mantle}.{\BBCQ}
\newblock
\APACjournalVolNumPages{Geophysical Journal International}{151}{1}{88--105}.
\newblock
\begin{APACrefDOI} \doi{10.1046/j.1365-246X.2002.01742.x} \end{APACrefDOI}
\PrintBackRefs{\CurrentBib}

\bibitem [\protect \citeauthoryear {%
Shen%
\ \protect \BOthers {.}}{%
Shen%
\ \protect \BOthers {.}}{%
{\protect \APACyear {2016}}%
}]{%
shen_2016_Seismic}
\APACinsertmetastar {%
shen_2016_Seismic}%
\begin{APACrefauthors}%
Shen, W.%
, Ritzwoller, M\BPBI H.%
, Kang, D.%
, Kim, Y.%
, Lin, F\BHBI C.%
, Ning, J.%
\BDBL {}Zhou, L.%
\end{APACrefauthors}%
\unskip\
\newblock
\APACrefYearMonthDay{2016}{}{}.
\newblock
{\BBOQ}\APACrefatitle {A Seismic Reference Model for the Crust and Uppermost Mantle beneath China from Surface Wave Dispersion} {A seismic reference model for the crust and uppermost mantle beneath china from surface wave dispersion}.{\BBCQ}
\newblock
\APACjournalVolNumPages{Geophysical Journal International}{206}{2}{954--979}.
\newblock
\begin{APACrefDOI} \doi{10.1093/gji/ggw175} \end{APACrefDOI}
\PrintBackRefs{\CurrentBib}

\bibitem [\protect \citeauthoryear {%
Shen%
, Ritzwoller%
, {Schulte-Pelkum}%
\BCBL {}\ \BBA {} Lin%
}{%
Shen%
, Ritzwoller%
, {Schulte-Pelkum}%
\BCBL {}\ \BBA {} Lin%
}{%
{\protect \APACyear {2013}}%
}]{%
shen_2013_Joint}
\APACinsertmetastar {%
shen_2013_Joint}%
\begin{APACrefauthors}%
Shen, W.%
, Ritzwoller, M\BPBI H.%
, {Schulte-Pelkum}, V.%
\BCBL {}\ \BBA {} Lin, F\BHBI C.%
\end{APACrefauthors}%
\unskip\
\newblock
\APACrefYearMonthDay{2013}{}{}.
\newblock
{\BBOQ}\APACrefatitle {Joint Inversion of Surface Wave Dispersion and Receiver Functions: A Bayesian Monte-Carlo Approach} {Joint inversion of surface wave dispersion and receiver functions: A bayesian monte-carlo approach}.{\BBCQ}
\newblock
\APACjournalVolNumPages{Geophysical Journal International}{192}{2}{807--836}.
\newblock
\begin{APACrefDOI} \doi{10.1093/gji/ggs050} \end{APACrefDOI}
\PrintBackRefs{\CurrentBib}

\bibitem [\protect \citeauthoryear {%
Shen%
, Ritzwoller%
\BCBL {}\ \BBA {} Schulte-Pelkum%
}{%
Shen%
, Ritzwoller%
\BCBL {}\ \BBA {} Schulte-Pelkum%
}{%
{\protect \APACyear {2013}}%
}]{%
shen_2013_3D}
\APACinsertmetastar {%
shen_2013_3D}%
\begin{APACrefauthors}%
Shen, W.%
, Ritzwoller, M\BPBI H.%
\BCBL {}\ \BBA {} Schulte-Pelkum, V.%
\end{APACrefauthors}%
\unskip\
\newblock
\APACrefYearMonthDay{2013}{}{}.
\newblock
{\BBOQ}\APACrefatitle {A 3-D Model of the Crust and Uppermost Mantle beneath the Central and Western US by Joint Inversion of Receiver Functions and Surface Wave Dispersion} {A 3-d model of the crust and uppermost mantle beneath the central and western us by joint inversion of receiver functions and surface wave dispersion}.{\BBCQ}
\newblock
\APACjournalVolNumPages{Journal of Geophysical Research: Solid Earth}{118}{1}{262--276}.
\newblock
\begin{APACrefDOI} \doi{10.1029/2012JB009602} \end{APACrefDOI}
\PrintBackRefs{\CurrentBib}

\bibitem [\protect \citeauthoryear {%
L.~Socco%
\ \BBA {} Strobbia%
}{%
L.~Socco%
\ \BBA {} Strobbia%
}{%
{\protect \APACyear {2004}}%
}]{%
socco_2004_Surfacewave}
\APACinsertmetastar {%
socco_2004_Surfacewave}%
\begin{APACrefauthors}%
Socco, L.%
\BCBT {}\ \BBA {} Strobbia, C.%
\end{APACrefauthors}%
\unskip\
\newblock
\APACrefYearMonthDay{2004}{}{}.
\newblock
{\BBOQ}\APACrefatitle {Surface-wave Method for Near-surface Characterization: A Tutorial} {Surface-wave method for near-surface characterization: A tutorial}.{\BBCQ}
\newblock
\APACjournalVolNumPages{Near Surface Geophysics}{2}{4}{165--185}.
\newblock
\begin{APACrefDOI} \doi{10.3997/1873-0604.2004015} \end{APACrefDOI}
\PrintBackRefs{\CurrentBib}

\bibitem [\protect \citeauthoryear {%
L\BPBI V.~Socco%
\ \BBA {} Boiero%
}{%
L\BPBI V.~Socco%
\ \BBA {} Boiero%
}{%
{\protect \APACyear {2008}}%
}]{%
socco_2008_Improved}
\APACinsertmetastar {%
socco_2008_Improved}%
\begin{APACrefauthors}%
Socco, L\BPBI V.%
\BCBT {}\ \BBA {} Boiero, D.%
\end{APACrefauthors}%
\unskip\
\newblock
\APACrefYearMonthDay{2008}{}{}.
\newblock
{\BBOQ}\APACrefatitle {Improved Monte Carlo Inversion of Surface Wave Data} {Improved monte carlo inversion of surface wave data}.{\BBCQ}
\newblock
\APACjournalVolNumPages{Geophysical Prospecting}{56}{3}{357--371}.
\newblock
\begin{APACrefDOI} \doi{10.1111/j.1365-2478.2007.00678.x} \end{APACrefDOI}
\PrintBackRefs{\CurrentBib}

\bibitem [\protect \citeauthoryear {%
L\BPBI V.~Socco%
, Foti%
\BCBL {}\ \BBA {} Boiero%
}{%
L\BPBI V.~Socco%
\ \protect \BOthers {.}}{%
{\protect \APACyear {2010}}%
}]{%
socco_2010_Surfacewave}
\APACinsertmetastar {%
socco_2010_Surfacewave}%
\begin{APACrefauthors}%
Socco, L\BPBI V.%
, Foti, S.%
\BCBL {}\ \BBA {} Boiero, D.%
\end{APACrefauthors}%
\unskip\
\newblock
\APACrefYearMonthDay{2010}{}{}.
\newblock
{\BBOQ}\APACrefatitle {Surface-Wave Analysis for Building near-Surface Velocity Models --- Established Approaches and New Perspectives} {Surface-wave analysis for building near-surface velocity models --- established approaches and new perspectives}.{\BBCQ}
\newblock
\APACjournalVolNumPages{Geophysics}{75}{5}{75A83-75A102}.
\newblock
\begin{APACrefDOI} \doi{10.1190/1.3479491} \end{APACrefDOI}
\PrintBackRefs{\CurrentBib}

\bibitem [\protect \citeauthoryear {%
Song%
, Tang%
, Lv%
, Fang%
\BCBL {}\ \BBA {} Gu%
}{%
Song%
\ \protect \BOthers {.}}{%
{\protect \APACyear {2012}}%
}]{%
song_2012_Application}
\APACinsertmetastar {%
song_2012_Application}%
\begin{APACrefauthors}%
Song, X.%
, Tang, L.%
, Lv, X.%
, Fang, H.%
\BCBL {}\ \BBA {} Gu, H.%
\end{APACrefauthors}%
\unskip\
\newblock
\APACrefYearMonthDay{2012}{}{}.
\newblock
{\BBOQ}\APACrefatitle {Application of Particle Swarm Optimization to Interpret Rayleigh Wave Dispersion Curves} {Application of particle swarm optimization to interpret rayleigh wave dispersion curves}.{\BBCQ}
\newblock
\APACjournalVolNumPages{Journal of Applied Geophysics}{84}{}{1--13}.
\newblock
\begin{APACrefDOI} \doi{10.1016/j.jappgeo.2012.05.011} \end{APACrefDOI}
\PrintBackRefs{\CurrentBib}

\bibitem [\protect \citeauthoryear {%
Tarantola%
}{%
Tarantola%
}{%
{\protect \APACyear {1984}}%
}]{%
tarantola_1984_Linearized}
\APACinsertmetastar {%
tarantola_1984_Linearized}%
\begin{APACrefauthors}%
Tarantola, A.%
\end{APACrefauthors}%
\unskip\
\newblock
\APACrefYearMonthDay{1984}{}{}.
\newblock
{\BBOQ}\APACrefatitle {Linearized Inversion of Seismic Reflection Data} {Linearized inversion of seismic reflection data}.{\BBCQ}
\newblock
\APACjournalVolNumPages{Geophysical Prospecting}{32}{6}{998--1015}.
\newblock
\begin{APACrefDOI} \doi{10.1111/j.1365-2478.1984.tb00751.x} \end{APACrefDOI}
\PrintBackRefs{\CurrentBib}

\bibitem [\protect \citeauthoryear {%
Thomson%
}{%
Thomson%
}{%
{\protect \APACyear {1950}}%
}]{%
thomson_1950_Transmission}
\APACinsertmetastar {%
thomson_1950_Transmission}%
\begin{APACrefauthors}%
Thomson, W\BPBI T.%
\end{APACrefauthors}%
\unskip\
\newblock
\APACrefYearMonthDay{1950}{}{}.
\newblock
{\BBOQ}\APACrefatitle {Transmission of Elastic Waves through a Stratified Solid Medium} {Transmission of elastic waves through a stratified solid medium}.{\BBCQ}
\newblock
\APACjournalVolNumPages{Journal of Applied Physics}{21}{2}{89--93}.
\newblock
\begin{APACrefDOI} \doi{10.1063/1.1699629} \end{APACrefDOI}
\PrintBackRefs{\CurrentBib}

\bibitem [\protect \citeauthoryear {%
Wang%
, Song%
\BCBL {}\ \BBA {} Li%
}{%
Wang%
\ \protect \BOthers {.}}{%
{\protect \APACyear {2023}}%
}]{%
wang_2023_Deeplearningbased}
\APACinsertmetastar {%
wang_2023_Deeplearningbased}%
\begin{APACrefauthors}%
Wang, F.%
, Song, X.%
\BCBL {}\ \BBA {} Li, M.%
\end{APACrefauthors}%
\unskip\
\newblock
\APACrefYearMonthDay{2023}{}{}.
\newblock
{\BBOQ}\APACrefatitle {A Deep-Learning-Based Approach for Seismic Surface-Wave Dispersion Inversion (SfNet) with Application to the Chinese Mainland} {A deep-learning-based approach for seismic surface-wave dispersion inversion (sfnet) with application to the chinese mainland}.{\BBCQ}
\newblock
\APACjournalVolNumPages{Earthquake Science}{36}{2}{147--168}.
\newblock
\begin{APACrefDOI} \doi{10.1016/j.eqs.2023.02.007} \end{APACrefDOI}
\PrintBackRefs{\CurrentBib}

\bibitem [\protect \citeauthoryear {%
Xia%
, Miller%
\BCBL {}\ \BBA {} Park%
}{%
Xia%
\ \protect \BOthers {.}}{%
{\protect \APACyear {1999}}%
}]{%
xia_1999_Estimation}
\APACinsertmetastar {%
xia_1999_Estimation}%
\begin{APACrefauthors}%
Xia, J.%
, Miller, R\BPBI D.%
\BCBL {}\ \BBA {} Park, C\BPBI B.%
\end{APACrefauthors}%
\unskip\
\newblock
\APACrefYearMonthDay{1999}{}{}.
\newblock
{\BBOQ}\APACrefatitle {Estimation of Near-surface Shear-wave Velocity by Inversion of Rayleigh Waves} {Estimation of near-surface shear-wave velocity by inversion of rayleigh waves}.{\BBCQ}
\newblock
\APACjournalVolNumPages{Geophysics}{64}{3}{691--700}.
\newblock
\begin{APACrefDOI} \doi{10.1190/1.1444578} \end{APACrefDOI}
\PrintBackRefs{\CurrentBib}

\bibitem [\protect \citeauthoryear {%
Xiao%
\ \protect \BOthers {.}}{%
Xiao%
\ \protect \BOthers {.}}{%
{\protect \APACyear {2024}}%
}]{%
xiao_2024_CSRM10}
\APACinsertmetastar {%
xiao_2024_CSRM10}%
\begin{APACrefauthors}%
Xiao, X.%
, Cheng, S.%
, Wu, J.%
, Wang, W.%
, Sun, L.%
, Wang, X.%
\BDBL {}Wen, L.%
\end{APACrefauthors}%
\unskip\
\newblock
\APACrefYearMonthDay{2024}{}{}.
\newblock
{\BBOQ}\APACrefatitle {CSRM-1.0: A China Seismological Reference Model} {Csrm-1.0: A china seismological reference model}.{\BBCQ}
\newblock
\APACjournalVolNumPages{Journal of Geophysical Research: Solid Earth}{129}{9}{e2024JB029520}.
\newblock
\begin{APACrefDOI} \doi{10.1029/2024JB029520} \end{APACrefDOI}
\PrintBackRefs{\CurrentBib}

\bibitem [\protect \citeauthoryear {%
Xiao%
, Sun%
, Wang%
\BCBL {}\ \BBA {} Wen%
}{%
Xiao%
\ \protect \BOthers {.}}{%
{\protect \APACyear {2022}}%
}]{%
xiao_2022_Simultaneous}
\APACinsertmetastar {%
xiao_2022_Simultaneous}%
\begin{APACrefauthors}%
Xiao, X.%
, Sun, L.%
, Wang, X.%
\BCBL {}\ \BBA {} Wen, L.%
\end{APACrefauthors}%
\unskip\
\newblock
\APACrefYearMonthDay{2022}{}{}.
\newblock
{\BBOQ}\APACrefatitle {Simultaneous Inversion for Surface Wave Phase Velocity and Earthquake Centroid Parameters: Methodology and Application} {Simultaneous inversion for surface wave phase velocity and earthquake centroid parameters: Methodology and application}.{\BBCQ}
\newblock
\APACjournalVolNumPages{Journal of Geophysical Research: Solid Earth}{127}{9}{e2022JB024018}.
\newblock
\begin{APACrefDOI} \doi{10.1029/2022JB024018} \end{APACrefDOI}
\PrintBackRefs{\CurrentBib}

\bibitem [\protect \citeauthoryear {%
F.~Yang%
\ \BBA {} Ma%
}{%
F.~Yang%
\ \BBA {} Ma%
}{%
{\protect \APACyear {2019}}%
}]{%
yang_2019_Deeplearning}
\APACinsertmetastar {%
yang_2019_Deeplearning}%
\begin{APACrefauthors}%
Yang, F.%
\BCBT {}\ \BBA {} Ma, J.%
\end{APACrefauthors}%
\unskip\
\newblock
\APACrefYearMonthDay{2019}{}{}.
\newblock
\APACrefbtitle {Deep-Learning Inversion: A next Generation Seismic Velocity-Model Building Method} {Deep-learning inversion: A next generation seismic velocity-model building method}\ (\BNUM\ arXiv:1902.06267).
\newblock
\APACaddressPublisher{}{arXiv}.
\newblock
\begin{APACrefDOI} \doi{10.48550/arXiv.1902.06267} \end{APACrefDOI}
\PrintBackRefs{\CurrentBib}

\bibitem [\protect \citeauthoryear {%
J.~Yang%
, Xu%
\BCBL {}\ \BBA {} Zhang%
}{%
J.~Yang%
\ \protect \BOthers {.}}{%
{\protect \APACyear {2022}}%
}]{%
yang_2022_Reconstruction}
\APACinsertmetastar {%
yang_2022_Reconstruction}%
\begin{APACrefauthors}%
Yang, J.%
, Xu, C.%
\BCBL {}\ \BBA {} Zhang, Y.%
\end{APACrefauthors}%
\unskip\
\newblock
\APACrefYearMonthDay{2022}{}{}.
\newblock
{\BBOQ}\APACrefatitle {Reconstruction of the S-Wave Velocity via Mixture Density Networks With a New Rayleigh Wave Dispersion Function} {Reconstruction of the s-wave velocity via mixture density networks with a new rayleigh wave dispersion function}.{\BBCQ}
\newblock
\APACjournalVolNumPages{IEEE Transactions on Geoscience and Remote Sensing}{60}{}{1--13}.
\newblock
\begin{APACrefDOI} \doi{10.1109/TGRS.2022.3169236} \end{APACrefDOI}
\PrintBackRefs{\CurrentBib}

\bibitem [\protect \citeauthoryear {%
Y.~Yang%
\ \BBA {} Forsyth%
}{%
Y.~Yang%
\ \BBA {} Forsyth%
}{%
{\protect \APACyear {2006}}%
}]{%
yang_2006_Regional}
\APACinsertmetastar {%
yang_2006_Regional}%
\begin{APACrefauthors}%
Yang, Y.%
\BCBT {}\ \BBA {} Forsyth, D\BPBI W.%
\end{APACrefauthors}%
\unskip\
\newblock
\APACrefYearMonthDay{2006}{}{}.
\newblock
{\BBOQ}\APACrefatitle {Regional Tomographic Inversion of the Amplitude and Phase of Rayleigh Waves with 2-D Sensitivity Kernels} {Regional tomographic inversion of the amplitude and phase of rayleigh waves with 2-d sensitivity kernels}.{\BBCQ}
\newblock
\APACjournalVolNumPages{Geophysical Journal International}{166}{3}{1148--1160}.
\newblock
\begin{APACrefDOI} \doi{10.1111/j.1365-246X.2006.02972.x} \end{APACrefDOI}
\PrintBackRefs{\CurrentBib}

\bibitem [\protect \citeauthoryear {%
Yao%
, Van Der~Hilst%
\BCBL {}\ \BBA {} De~Hoop%
}{%
Yao%
\ \protect \BOthers {.}}{%
{\protect \APACyear {2006}}%
}]{%
yao_2006_Surfacewave}
\APACinsertmetastar {%
yao_2006_Surfacewave}%
\begin{APACrefauthors}%
Yao, H.%
, Van Der~Hilst, R\BPBI D.%
\BCBL {}\ \BBA {} De~Hoop, M\BPBI V.%
\end{APACrefauthors}%
\unskip\
\newblock
\APACrefYearMonthDay{2006}{}{}.
\newblock
{\BBOQ}\APACrefatitle {Surface-Wave Array Tomography in SE Tibet from Ambient Seismic Noise and Two-Station Analysis - I. Phase Velocity Maps} {Surface-wave array tomography in se tibet from ambient seismic noise and two-station analysis - i. phase velocity maps}.{\BBCQ}
\newblock
\APACjournalVolNumPages{Geophysical Journal International}{166}{2}{732--744}.
\newblock
\begin{APACrefDOI} \doi{10.1111/j.1365-246X.2006.03028.x} \end{APACrefDOI}
\PrintBackRefs{\CurrentBib}

\bibitem [\protect \citeauthoryear {%
Zhang%
, Zhang%
, Liu%
\BCBL {}\ \BBA {} Fan%
}{%
Zhang%
\ \protect \BOthers {.}}{%
{\protect \APACyear {2022}}%
}]{%
zhang_2022_Deep}
\APACinsertmetastar {%
zhang_2022_Deep}%
\begin{APACrefauthors}%
Zhang, L.%
, Zhang, G.%
, Liu, Y.%
\BCBL {}\ \BBA {} Fan, Z.%
\end{APACrefauthors}%
\unskip\
\newblock
\APACrefYearMonthDay{2022}{}{}.
\newblock
{\BBOQ}\APACrefatitle {Deep Learning for 3-D Inversion of Gravity Data} {Deep learning for 3-d inversion of gravity data}.{\BBCQ}
\newblock
\APACjournalVolNumPages{IEEE Transactions on Geoscience and Remote Sensing}{60}{}{1--18}.
\newblock
\begin{APACrefDOI} \doi{10.1109/TGRS.2021.3110606} \end{APACrefDOI}
\PrintBackRefs{\CurrentBib}

\end{thebibliography}
%TC:endignore

\end{document}